\documentclass[10pt,prd,preprintnumbers,floatfix,aps,nofootinbib,showpacs,twocolumn,superscriptaddress,noshowpacs]{revtex4-2} 

\usepackage{aas_macros}
\usepackage{epsfig}
\usepackage{url}
\usepackage{hyperref}

\usepackage[normalem]{ulem} 

\usepackage{latexsym}
\usepackage{epsfig}
\usepackage{amsmath}
\usepackage{amssymb}
\usepackage{wasysym}
\usepackage{graphicx}
\usepackage{dcolumn}
\usepackage{verbatim}
\usepackage{enumerate,mdwlist}
\usepackage[titletoc]{appendix}
\usepackage{amsfonts}
\usepackage{fancyvrb} 
\usepackage{tikz} 
\usetikzlibrary{calc}
\usepackage[export]{adjustbox}

\usepackage{booktabs}
\usepackage{siunitx}   

\usepackage[normalem]{ulem}

\newcommand{\ba}{\begin{eqnarray}}
\newcommand{\ea}{\end{eqnarray}}
\newcommand{\be}{\begin{equation}}
\newcommand{\ee}{\end{equation}}

\newcommand{\mur}{$\mu_{r} \ $}
\newcommand{\delt}{$\Delta T \ $}
\newcommand{\murdelt}{$\mu_r-\Delta T \ $}

\newcommand{\glow}{\texttt{GLoW} \ }

\definecolor{grey}{rgb}{0.4,0.4,0.4}
\definecolor{dullmagenta}{rgb}{0.4,0,0.4}
\definecolor{darkblue}{rgb}{0,0,0.4}
\definecolor{midblue}{rgb}{0,0,0.5}
\definecolor{midred}{rgb}{0.5,0,0}
\definecolor{orange}{rgb}{1,0.5,0}
\definecolor{lightbrown}{rgb}{0.75,0.5,0.25}
\definecolor{tan}{cmyk}{0.14,0.42,0.56,0}
\definecolor{djunglegreen}{cmyk}{0.99,0,0.52,0}
\definecolor{lightgreen}{rgb}{0,1,0}
\definecolor{olivegreen}{cmyk}{0.64,0,0.95,0.40}
\definecolor{midgreen}{rgb}{0.0,0.675,0.0}
\definecolor{darkgreen}{rgb}{0,0.5,0}
\definecolor{ceruleanblue}{rgb}{0.0, 0.2, 0.7}
\definecolor{burgundy}{rgb}{0.5, 0.0, 0.13}
\definecolor{oxblood}{rgb}{0.5333, 0.0314, 0.0314}

\definecolor{hvred}{RGB}{186,12,47}

\hypersetup{
    colorlinks=true,
    linkcolor=oxblood,
    filecolor=oxblood,      
    urlcolor=oxblood,
    citecolor=oxblood,
}

\usepackage[all]{xy} 
\usepackage{amsfonts}
\providecommand{\selectlanguage}[1]{}

\makeatletter
\def\l@subsubsection#1#2{}
\makeatother

\begin{document}

\title{Dark Matter Subhalos and Higher Order Catastrophes in Gravitational Wave Lensing}

\author{Luka Vujeva}
\email{luka.vujeva@nbi.ku.dk}
\affiliation{Center of Gravity, Niels Bohr Institute, Blegdamsvej 17, 2100 Copenhagen, Denmark}
\author{Jose Mar\'ia Ezquiaga}
\affiliation{Center of Gravity, Niels Bohr Institute, Blegdamsvej 17, 2100 Copenhagen, Denmark}
\author{Daniel Gilman}
\affiliation{Department of Astronomy \& Astrophysics, University of Chicago, Chicago, IL 60637, USA}

\author{Srashti Goyal}
\affiliation{Max Planck Institute for Gravitational Physics (Albert Einstein Institute)\\
Am Mühlenberg 1, D-14476 Potsdam-Golm, Germany}
\author{Miguel Zumalac\'arregui}
\affiliation{Max Planck Institute for Gravitational Physics (Albert Einstein Institute)\\
Am Mühlenberg 1, D-14476 Potsdam-Golm, Germany}

\begin{abstract}

Gravitational lensing is an invaluable probe of the nature of dark matter, and the structures it forms.
Lensed gravitational waves 
in particular 
allow for unparalleled sensitivity to small scale structures within the lenses, due to the precise time resolution 
in combination with the continuous monitoring of the entire sky. 
In this work, we show two distinct ways of using strongly lensed gravitational waves to identify the presence of dark matter subhalos: 
\emph{{i)}} through higher order caustics generating high relative magnification ($\mu_r > 2$), short time delay image pairs that break the caustic universality relations of single dark matter halos, which occur for $\sim 1-10$ percent of strongly lensed events in our cold dark matter models,
and \emph{ii)}
through the presence of more than three highly magnified images, which occur for $\sim 0.01-1$ percent of the same simulated events.
We find that these results are highly sensitive to the concentrations of subhalos in our simulations, and more mildly to their number densities.
The presence of low-mass subhalos increases the probability of observing wave-optics lensing 
in lensed gravitational waves, which is studied by solving the diffraction integral 
with the stationary phase approximation, as well as numerically. 
We also report distinct quantitative and qualitative differences in the distributions of relative magnifications and time delays for subhalo populations with increased number densities or concentrations.
With the upcoming detection
of strongly lensed events  
by ground- and space-
based detectors, comparisons against these simulated distributions will provide insight into the nature of dark matter.
\end{abstract}

\date{\today}

\maketitle


\section{Introduction}\label{sec:intro}

\begin{figure*}
    \centering
    \includegraphics[width=\linewidth]{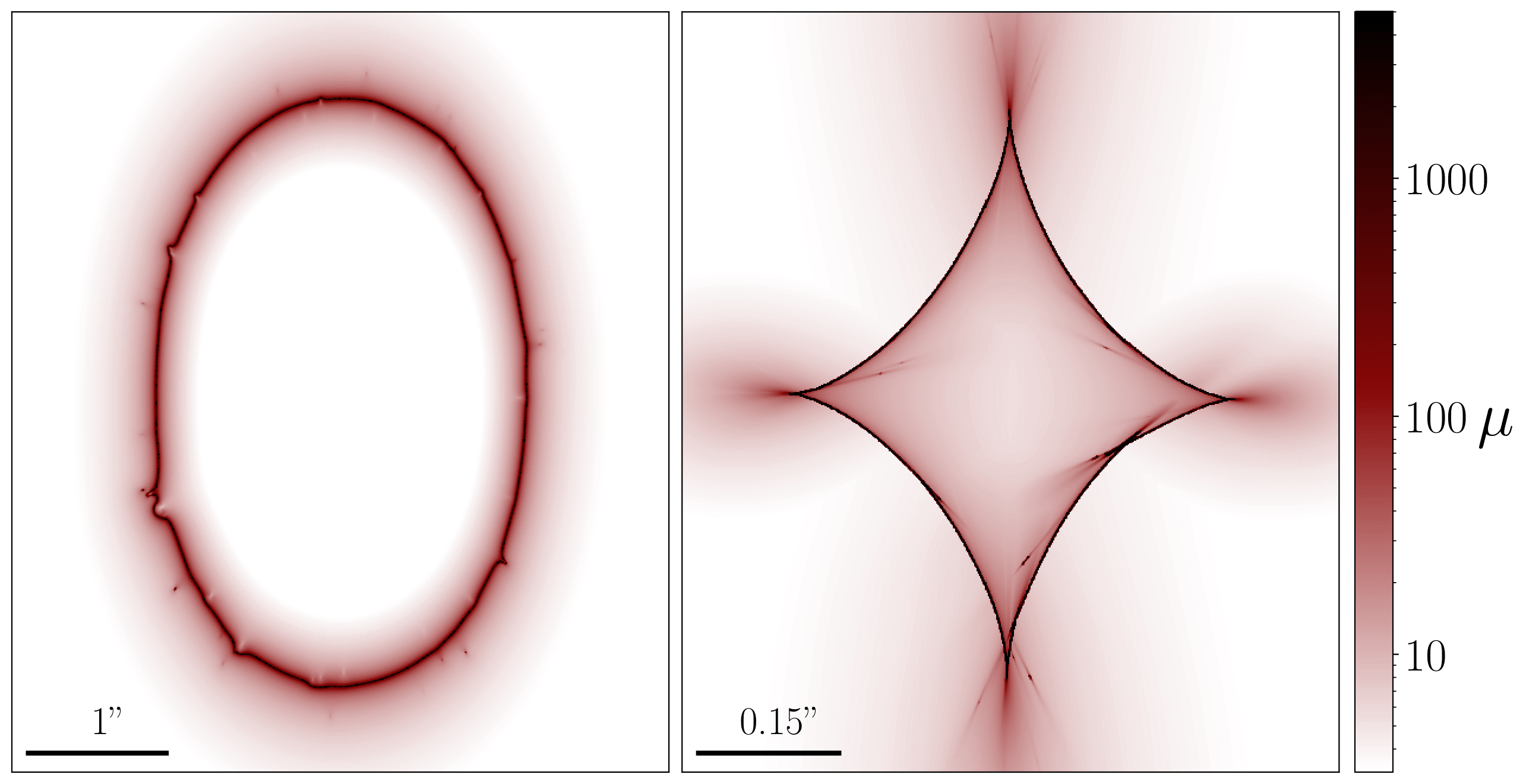}

    \caption{Magnification ($\mu$) map of an example galaxy populated with cold dark matter subhalos in the image (left) and source (right) plane. Note that this example corresponds to the increased concentration case explored later in the work.
    }
    \label{fig:mag_cusp_vs_fold_subhalos}
\end{figure*}

Gravitational lensing offers striking glimpses into the high-redshift Universe by nature of the 'gravitational telescope' phenomenon caused massive objects such as galaxies and galaxy clusters.
This has provided us access into previously inaccessible periods of the evolution of the Universe, such as allowing us to observe galaxies \cite{Wang:2023apjl, kokorev:2025apj} and even individual stars \cite{Diego:2022nzw,welch:2022nat} at unprecedented distances beyond the sensitivity of current telescopes \cite{Mason:2015cna, vujeva:2024sur}.
Additionally, it has allowed for the most stringent observational tests of the nature of dark matter, and the structures it forms \cite{Vegetti:2023mgp,Dutra:2024qac}. 
While there has been a tremendous amount of success in observing gravitational lensing in the electromagnetic (EM) spectrum (and much more to come with all sky observation from the Vera Rubin Observatory \cite{LSST:2008ijt} and the Euclid telescope \cite{Euclid:2021icp, Euclid:ER1}), focusing on different messengers can unveil new information about gravitational lensing \cite{Smith:2025axx}.

Gravitational waves (GWs) are unique transients because they are coherently detected with a timing precision down to the millisecond. 
This is to be contrasted with the current state-of-the-art in EM lensing, constraining differences in the arrival times of images of the order of days \cite{Millon:2020xab}. 
Moreover, their waveform morphology and phase evolution are accurately recorded with the LIGO \cite{LIGOScientific:2014pky}, VIRGO \cite{VIRGO:2014yos}, and KAGRA \cite{KAGRA:2020tym} (LVK) detectors, with remarkable recent examples~\cite{LIGOScientific:2025rid,LIGOScientific:2025cmm}. 
Therefore, lensed GWs are an exceptional new probe of the dark matter (DM) substructures, which intrinsically predict short lensing time delays.

Although we still have 
to detect a strongly lensed GWs \cite{Hannuksela:2019kle,LIGOScientific:2021izm,Janquart:2023mvf,LIGOScientific:2023bwz}, their discovery is expected in the next observing run~\cite{Ng:2017yiu,Oguri:2018muv,Xu:2021bfn,Wierda:2021upe,Smith:2022vbp}.
This will be further improved in future ground-based detectors such as Einstein Telescope~\cite{ET:2019dnz} and Cosmic Explorer~\cite{Reitze:2019iox},
and future space-based detectors such as LISA \cite{LISA:2017pwj, LISA:2024hlh}, which offer exciting prospects for measuring 
wave-optics phenomena from larger lenses such as galaxies, supermassive black holes or subhalos.

The focus of GW lensing is typically divided between the study of large scale structures such as galaxies and galaxy clusters \cite{Robertson:2020mfh,Dai:2018mxx,Oguri:2018muv, Oguri:2019fix, Lo:2024wqm, Vujeva:2025kko}, and the interference effects of compact objects such as stars and black holes \cite{Pijnenburg:2024btj,Chan:2024qmb,Chan:2025wgz}.
While recent works have explored GW lensing with subhalos in the single image regime \cite{Caliskan:2023zqm,Savastano:2023spl, Brando:2024inp}, this work aims to explore the strong lensing phenomenology of these systems, and potential interference effects near the consequent caustics, which may differ from the universal predictions without substructure~\cite{Lo:2024wqm, Ezquiaga:2025gkd}.

In the context of GW lensing from galaxy-scale lenses, most work considers single, spheroidal dark matter halos following typical dark matter distributions such as Singular Isothermal Spheres (SIS), Navarro-Frenk-White (NFW) \cite{Navarro:1996gj, Navarro:1997mt, Taylor:2001gk}, or Singular Isothermal Ellipsoids (SIE) profiles \cite{1993ApJ...417..450K, 1994A&A...284..285K, Barkana:1998qu}.
While these profiles are adequate to describe the main dark matter halo of galaxies, this work aims to be the first to consider the effects of dark matter subhalos on strongly lensed GWs in these traditionally smooth, singular lenses. 

The existence of DM subhalos is motivated by the hierarchical mergers seen in N-body cosmological simulations, which find that a fraction of the total dark matter mass in large DM halos is contained in lower mass, typically more concentrated DM subhalos \cite{Springel:2008cc,Fiacconi:2016cih,Diemer:2018vmz,Nadler:2022dvo}. 
While not observed directly, their existence has been put forward as a solution to some prevalent issues in EM lensing, such as quasar flux anomalies \cite{Schechter:2002dm,Gilman:2019nap,Nierenberg:2023tvi}, deviations from expected image locations \cite{Ephremidze:2025mqg}, and excess total magnification of sources near caustics \cite{Aazami:2006qw}.

Many alternative DM models create differences in the low end of the subhalo mass distribution (as well as the density distribution of DM halos in certain models). Cold DM (CDM) allows for the creation of much smaller DM halos when compared to warm, fuzzy, wave, and self interacting DM, which typically suppress the production of structure on small scales, leading to lower number densities of low mass halos when compared to CDM \cite{Zavala:2019gpq}. Therefore, being sensitive to low mass halos is crucial in probing the nature of DM. 
Moreover, lighter subhalos do not hold baryons efficiently, and therefore are a more direct probe of DM theories~\cite{Bullock:2017xww}.

In this work, we show that not only are strongly lensed GWs powerful probes of dark matter subhalos dow to masses of $10^7 M_\odot$ (and can be extended to much lower masses), but we also show that there are concrete criteria that can be used to show that even a single strongly lensed GW event has been affected by a subhalo. These deviations come in the forms of either seeing 4 (or more) highly magnified gravitational waves coming in short succession (which has been previously explored in the context of EM lensing \cite{Ji:2024ott}), or seeing a deviation in the time delay and relative magnifications of the two brightest images from the predictions coming from catastrophe theory for single smooth potentials. More specifically, we see that the higher order caustics (or catastrophes) caused by the inclusion of substructure in the form of subhalos can cause the brightest pair of images to have relative magnification factors of $\mu_r > 2$ at short time delays (\delt), which is impossible in a smooth single potential. The detection of such an event would be a signature of the presence subhalos (or other compact objects) effecting the lensed signals.

We can explicitly show that subhalos affect lensed images by the deviation from the expected \murdelt relation for short time delays and highly magnified GWs. Namely, without the subhalos, the two brightest images must have $\mu_r < 2$ at small $\Delta T$ (which is the upper limit set by the cusp caustic, with the behavior of images near the fold being $\mu_r \rightarrow 1$ at low $\Delta T$ \cite{Schneider:1992,Lo:2024wqm,Ezquiaga:2025gkd}). While deviations from the fold behavior are harder to detect (but could be studied from a large population of lensed GWs), subhalos near the images produced by cusps are highly sensitive to the subhalo perturber, and the production of an image pair with short time delay and $\mu_r > 2$ is smoking gun evidence of the existence of subhalos near caustics. In this work, we do not consider the impact of line-of-sight halos, which could also contribute to a similar effect \cite{Despali:2017ksx, Gilman:2019vca}.

This work is organized as follows: in Sec.~\ref{sec:lensing} we give an overview of gravitational lensing formalism in both the wave optics and geometric optics regimes, and introduce the formalism of typical caustics. In Sec.~\ref{sec:subhalos} we summarize our treatment of dark matter subhalos in our composite lens model. 
Sec.~\ref{sec:statistics} provides the results of our fiducial model, as well as exploring the impacts of modifying the dark matter model.
Finally, we examine the links back to higher order catastrophes in  Sec.~\ref{sec:results}. Throughout this work we adopt a Planck 2018 cosmology \cite{Planck:2018vyg}.

\section{Gravitational Lensing}\label{sec:lensing}

\begin{figure*}[t]
    \centering
    \includegraphics[width=\linewidth]{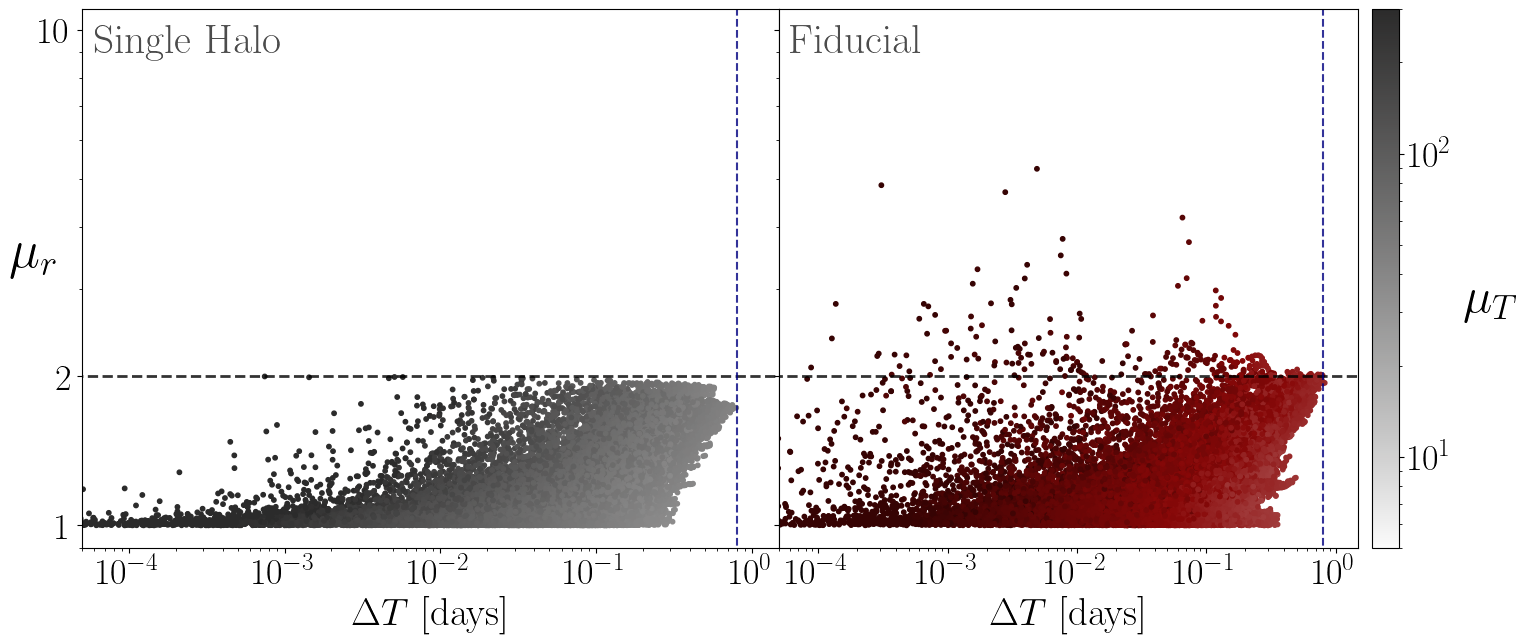}

    \caption{Relative magnification ($\mu_r$) vs time delay ($\Delta T$) for the two brightest images of sources placed near the caustics of a single elliptical single isothermal sphere halo without (left), and 
    with subhalos (right).  
    The grey lines in both plots represent the asymptotic values for the relative magnifications for sources very close to the caustics, where $\mu_r = 2$ is the limit for the cusp.
    Note that in the case of subhalo perturbers, the relative magnifications at low time delays can greatly exceed the theoretical limit set by the cusp caustic. The vertical blue line corresponds to the shortest time delay measured in the EM,a quadruply lensed quasar found to have a time delay of $\Delta T = 0.8 ^{+0.8}_{-0.7}$ days \cite{Millon:2020xab}.
    }
    \label{fig:td_mur_clean_vs_sub}
\end{figure*}

In this section, we review the fundamentals of gravitational lensing in both the wave and geometric optics regimes, and introduce the typical caustics seen in single dark matter halos, namely the fold and cusp caustics. 

\subsection{Wave Optics}\label{subsec:wo}

The changes in the amplitude of a lensed gravitational wave is characterized by the \textit{amplification factor} $F(f)$, simply defined as 

\begin{equation}
    F(f) = \frac{ \tilde h_L(f)}{\tilde h_0(f)},
\end{equation}\label{eq:F_f}

where $\tilde h_L(f)$ and $\tilde h_0(f)$ are the Fourier transformed lensed and unlensed gravitational wave strains.
The strain of the lensed gravitational wave is determined by the Kirchhoff diffraction integral \cite{Schneider:1992, Takahashi:2003ix, Tambalo:2022wlm}, given by 

\begin{equation}
    F(\omega) =  \frac{\omega}{2\pi i} \int d^2 \vec{x} e^{i\omega T(\vec{x}, \vec{y})},
\end{equation}\label{eq:F_f}

where $T(\vec{x}, \vec{y})$ is the \textit{Fermat potential}, and $\omega$ is the dimensionless frequency defined as $\omega \equiv 8 \pi G M_{Lz} f$.The redshifted lens mass $M_{Lz}$ is defines as

\begin{equation}
    M_{Lz} \equiv (1+z_L)\frac{D_S}{D_L D_{LS}}  \frac{\theta_E^2}{4G}.
\end{equation}

In units where image plane ($\vec{x}$) and source plane position are rescaled by a characteristic length scale in the system (typically the Einstein radius $\theta_E$) as $\theta = \vec{x}/\theta_E\, \ \beta = \vec{y}/\theta_E\,$, the Fermat potential, also called the \textit{time delay surface}~\cite{Blandford:1986zz, Schneider:1992} can be written as

\begin{equation}
    T_d(\theta,\beta) = \phi(\theta, \beta) \equiv  \bigg[  \frac{(\theta - \beta)^2}{2}  - \psi(\theta) \bigg ],
\end{equation}

where $\psi(\theta$) is the lens potential. The amplification factor can also be expressed in the time domain (which is what the lensing code \texttt{GLoW}~\cite{Villarrubia-Rojo:2024xcj} that is being used in this work solves for) through a simple Fourier transform of $F(\omega)$ 

\begin{equation}
I(\tau) = \int_{-\infty}^{\infty} \frac{iF(\omega)}{\omega} \, e^{-i \omega t} \, d\omega.
\end{equation}

Using this, we can also define the lensing Green's function, which is simply given by

\begin{equation}
    G(\tau) = \frac{1}{2\pi}\frac{d}{d\tau }I(\tau).
\end{equation}

The physical meaning of $G(\tau)$ is that its convolution with the unlensed signal gives us the desired lensed signal $h_L$.

\subsection{Geometric Optics}\label{subsec:go}

The geometric optics (GO) limit is valid for lensing systems in which the wavelength of the GW is much smaller than the characteristic size of the lens, and when a source of finite frequency is sufficiently far from a caustic.
 In this limit, GW lensing closely follows that of light \citep{Takahashi:2003ix}. 
Relating back to the wave optics regime, the image locations in geometric optics are determined by the stationary points of the stationary points of the integrand.
 For a given lensing configuration in GO, the locations of images are determined by the lens equation,
\begin{equation}
    \beta = \theta - \alpha(\theta),
\end{equation}\label{eq:lensequation}
where $\beta$ is the source location, $\theta$ is the image location, and $\alpha(\theta)$ is the deflection angle, determined by the lensing potential $\psi(\theta)$ as $\alpha(\theta) = \nabla \psi(\theta)$. The dimensionless surface mass density, or convergence, is defined as,
\begin{equation}
    \kappa(\theta) = \frac{\Sigma(\theta)}{\Sigma_c},
\end{equation}
where $\Sigma(\theta)$ is the surface mass density of the lens, and $\Sigma_c$ is the critical surface mass density at the redshift of the lens,

\begin{equation}
    \Sigma_c = \frac{c^2 D_S}{4 \pi G D_L D_{LS}},
\end{equation}
where $G$ is Newton's gravitational constant, $c$ is the speed of light, and $D_S$, $D_L$, and $D_{LS}$ are the angular diameter distances to the source, the lens, and between the lens and source respectively. The difference in arrival time between two images $\theta_i$ and $\theta_j$ of the same source at a position $\beta$ is
\begin{equation}
    \Delta T_{ij} = \frac{1+z_L}{c} \frac{D_L D_S}{D_{LS}} \big [ T_d(\theta_i, \beta) - T_d(\theta_j, \beta) \big ],
\end{equation}
where $z_L$ is the redshift of the lens

In this language, the lens equation and image positions $\vec\theta_i$ are simply given by $\left.\partial T_d /\partial \vec\theta \right|_{\vec\theta=\vec\theta_i}=0$. 
Finally, the magnification of a given image is defined as
\begin{equation}
    \mu^{-1} = (1-\kappa(\theta))^2 - \gamma(\theta)^2,
\end{equation}
where $\gamma(\theta)$ is the shear. 
This, again, can be derived directly from the time delay surface, i.e., $\mu^{-1}=\mathrm{det}\left(\partial^2T_d/\partial \vec\theta\partial\vec\theta\right)$. The points along which $|\mu| \rightarrow \infty$ in the image plane are called \emph{critical curves} (CCs), and their equivalent in the source plane are called \emph{caustics}. These will be explored in detail in the coming sections.

Given the highly oscillatory nature of the integrand in Eq. \ref{eq:F_f} at high frequencies, we employ the \textit{stationary phase approximation} (SPA) to produce lensed waveforms for sources whose frequencies would fall within the regime of ground based detectors such as the LVK network. The SPA assumes that most of the contribution to the integral comes from the stationary points of the integrand, which correspond to the image locations. This leads to the approximation (whose detailed derivation can be found in \cite{Schneider:1992, Ezquiaga:2020gdt}),
\begin{equation}
    F(w)\approx\sum_j \sqrt{|\mu(\vec x_j)|}\exp\left(iw T_d(\vec x_j) - i n_j \pi/2\right)\,,
\end{equation}
in which each image has an arrival time $T_j$, magnification $\mu_j$, and phase shift $n_j$. The validity of this approach for gravitational wave sources near caustics has been explored in previous works \cite{Ezquiaga:2025gkd, Serra:2025kbw}, and should be valid for the source locations considered in this work. Exploring methods of reliably calculating the full frequency dependent amplification factor at high dimensionless frequencies is left to future work. 

 A quantity that is often used throughout this work is the relative magnification factor, which (unless otherwise specified) is simply $\mu_r = |\mu_1 / \mu_2|$, where $\mu_1$ and $\mu_2$ are the brightest and second brightest images of a given source at a location $\beta$. We also define the parity of the images based on the sign of their magnification factors (i.e. a positive parity image has $\mu_i > 0$, whereas a negative parity image has $\mu_i < 0$).

\subsection{Fold and Cusp Caustics}\label{subsec:caustics}

In a non-axisymmetric lens, there are two types of caustics present in lensing configurations with a single potential: \textit{fold} and \textit{cusp} caustics. These have been explored in the context of gravitational waves for both the geometric and wave optics regimes in depth in Ezquiaga et al. \cite{Ezquiaga:2025gkd}, and point readers to this work for detailed derivations of the following expression. We summarize pertinent details in this sub-section, and how it will relate to the higher order caustics of interest in this work.

The properties of caustics (also known as catastrophes) are studied through an expansion of the time delay surface ($T_d$) around critical curves $\{\vec{x}\}$ and caustics $\{\vec{y}\}$. We choose our coordinate system such that the critical curves and caustics occur at $\vec{x} = 0$ and $\vec{y} = 0$ (i.e. the center) of the image plane and source plane respectively. We denote derivatives in the time delay surface $T_d$ as $T_{i_1\cdots i_{n}}$. Critical curves are found at the image positions $T_1 = T_2 = 0$ where the determinant of the Hessian matrix is
\begin{equation} \label{eq:detTab}
    D\equiv\mathrm{det}(T_{ab}) = T_{11}T_{22} - T_{12}^2=0\,.
\end{equation}  

The rank of the Hessian matrix $T_{ab}$ will determine the property of a generic caustic, where having a rank of 1 corresponds to the stable caustics \cite{Whitney1992} we are interested in, \textit{folds} and \textit{cusps}. A fold caustic is described by a lens mapping in which the determinant of the Hessian vanishes, it's normal vector does not vanish, and the Hessian itself does not vanish (i.e $T_{222} \neq 0$). In the case of the cusp, the condition are that $T_{11} \neq 0$ and $T_{222} = 0$, thus we would need to include quadratic derivatives in the $x_2$ direction, or in other words, $T_{2222} \neq 0$. A full derivation of the lensing quantities of sources near these caustics can be found in \cite{Ezquiaga:2025gkd}.

A fold caustic generated two images which can be highly magnified, whose magnifications are simply given by
\begin{equation}
    \mu_\pm=\pm\frac{1}{T_{11}\sqrt{2T_{222}y_2}}\,,
\end{equation}
 which shows that the relative magnification of a source in the universality regime near a fold caustic will always have $\mu_r = 1$. The arrival time difference of the images is 
\begin{equation}
    \Delta T =T_--T_+= \frac{4\sqrt{2}}{3}\frac{|y_2|^{3/2}}{|T_{222}|^{1/2}}\,,
\end{equation}
where one can also see that as $y_2 \rightarrow 0$, $\Delta T \rightarrow 0$,  leading to the rich phenomenology of interfering and diffractive signals.

 For the cusp, the equations become much more complicated, and for the purposes of this work we only show the expressions for a source approaching along the symmetry axis $y_2 = 0$. A source within a cusp can produce three highly magnified images. A source on the symmetry axis produces one one image of a given parity (dependent on the $\gamma$ parameter that is defined below) which arrives at the time

\begin{equation}
    T_d(\vec x^{(0)})=T_c-\frac{1}{2}\frac{y_1^2}{T_{11}}\,,
\end{equation}

, as well as two images of parities opposite to the first, which both arrive simultaneously at the time

\begin{equation}
    \Delta T = T_d(\vec x^{(1,2)})-T_d(\vec x^{(0)})= \frac{1}{2}\frac{T_{122}^2y_1^2}{T_{11}^2\gamma}\,.
\end{equation}

where $\gamma$ is defined as 

\begin{equation} \label{eq:gamma}
    \gamma\equiv \frac{T_{122}^2}{T_{11}} - \frac{T_{2222}}{3}. 
\end{equation}

The magnifications of these images are given by the expression

\begin{equation} \label{eq:magnification_cusp_symmetric}
    \mu^{(0)}=-2\mu^{(1,2)}=\frac{1}{T_{122}y_1}=\frac{T_{11}}{\sqrt{2\gamma\Delta T}}\,,
\end{equation}

which shows that the maximum relative magnification of these images is  $\mu_r = 2$.

\section{Composite Lens Model}\label{sec:subhalos}

\begin{figure}[t]
    \centering
    \includegraphics[width=\linewidth]{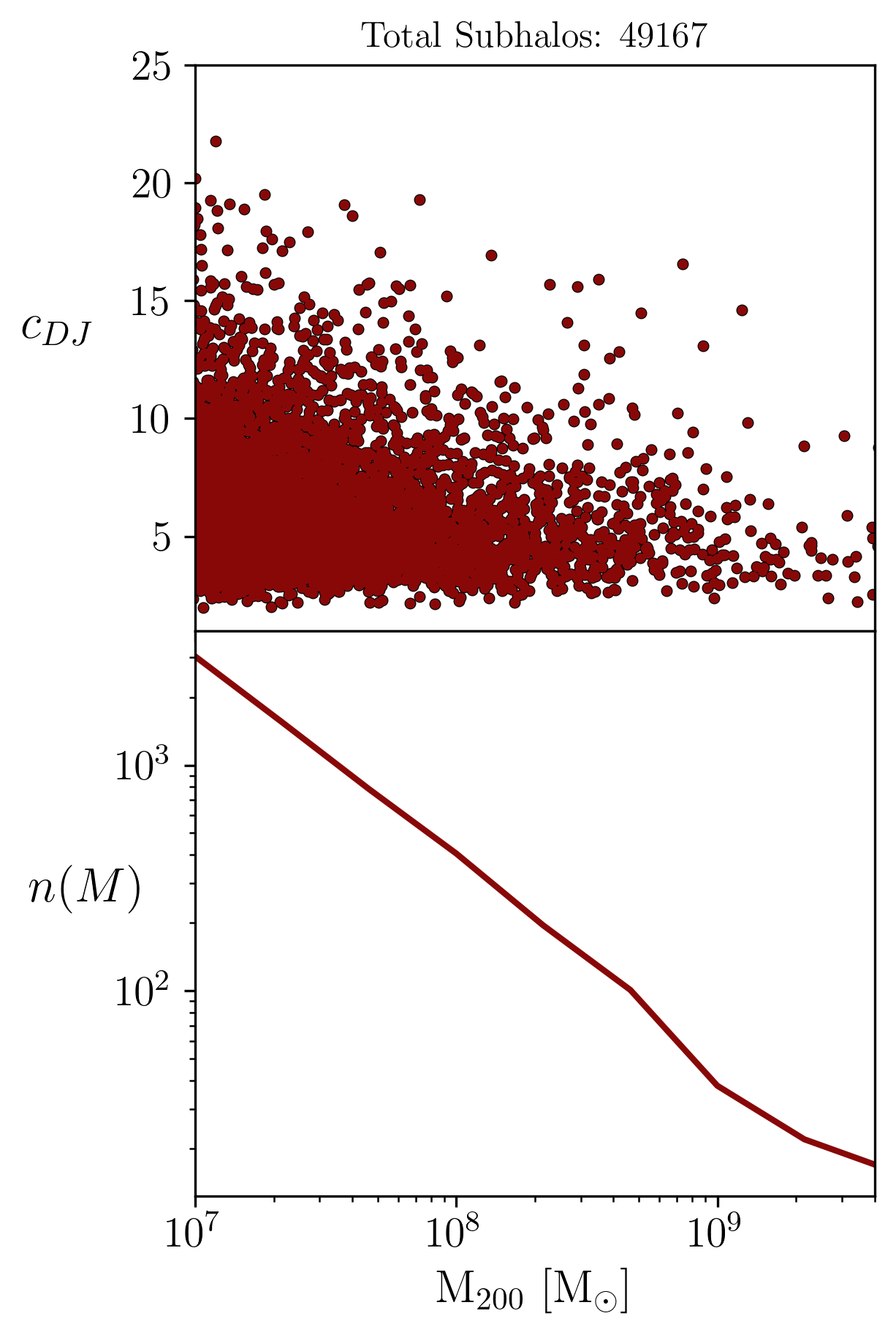}
    \caption{
    Concentrations (top) and number counts (bottom) of subhalos as a function of mass ($M_{200}$) for the fiducial Diemer-Joyce parameters~\cite{Diemer:2018vmz}, shown for an opening angle of 30 arcseconds.
    }
    \label{fig:number_con_mass}
\end{figure}

In this section, we outline the methods used to model dark matter subhalos in our lensing configuration. We explore the impact of DM subhalos on lensed gravitational waves by considering them to be embedded in a galaxy scale main halo.

The main halo used in this work is modeled using an elliptical single isothermal sphere (eSIS). The eSIS lensing potential is described as

\begin{equation}
    \psi(\vec{x}) = \sqrt{x_1^2 + x_2^2 / q^2},
\end{equation}

where $q$ controls the ellipticity of the halo, and $\psi_0$ is given by

\begin{equation}
    \psi_0 \equiv \frac{\sigma_v^2}{G \Sigma_{cr}\xi_0 },
\end{equation}

where $\sigma_v$ is the velocity dispersion of the halo.

To simulate the subhalo population, we use $\texttt{pyHalo}$\footnote{\url{https://github.com/dangilman/pyHalo}} \cite{Gilman:2019nap} to get the required subhalo properties such as their masses, concentrations, and radial distribution. Following Ref.~\cite{Gilman:2019nap}, the subhalo mass function is given by

\begin{equation}
    \frac{d^2N}{d\log mdA} = \Sigma_{sub}(m/m_0)^{-\alpha},
\end{equation}

where $\Sigma_{sub}$ is the normalization of the mass function at $10^8 M_\odot$, and $\alpha$ controls the logarithmic slope of the mass function. The subhalos considered in this work are all described with a Navarro-Frenk-White (NFW) profile, which has been shown to adequately describe CDM halos in $N-$body simulations \cite{Navarro:1996gj, Navarro:1997mt, Taylor:2001gk}. Note that we make the simplifying assumption of neglecting the tidal evolution models in \texttt{pyHalo}, and thus simply model the subhalos as NFW profiles to interface with \glow. Exploring the impact of including these effects is left to future studies.
They have density distributions that follow

\begin{equation}
    \rho (r) = \frac{\rho_s}{(r/r_s) (1 + r/r_s) ^2} \ ,
    \label{eq:nfwprofile}
\end{equation}

where $r_s$ is the scale radius, and $\rho_s/4$ is the density at $r_s$ (called the scale density). The size of the scale radius is controlled by the halo concentration $c$, through $r_s = r_{200}/c$, where $r_{200}$ is the radius at which the overdensity of the halo is two hundred times the critical density of the Universe at that redshift. The relationship between the mass and concentration of the subhalos (called the ``$c-M$'' relationship) is adopted from Diemer and Joyce \cite{Diemer:2018vmz} unless otherwise specified, and is implemented with a 0.2 dex scatter in their relationship \cite{Wang:2020hpl}. Concentrations described by this model will be denoted as $c_{DJ}$ onward. Varying the magnitude of this scatter was found to have a negligible overall impact on the results of this work.  

We model the combined smooth potential of the main deflector, including the stellar mass of the galaxy and its host dark matter halo, using an eSIS model with an Einstein radius of $\sim 9.4$ kpc (for at a lens plane redshift of $z_L = 0.5$, and source plane at $z_S =2$), and ellipticity of $q=0.9$, typical of galaxy-scale deflectors \cite{Auger:2010}.
The minimum subhalo mass considered in this work is $10^{-6} M_{main} = 10^7 M_\odot$. The subhalo mass function and related concentrations are shown in Fig.~\ref{fig:number_con_mass} for a line of sight with an opening angle of 30 arcseconds. 

\begin{figure}[t]
    \centering
    \includegraphics[width=\linewidth]{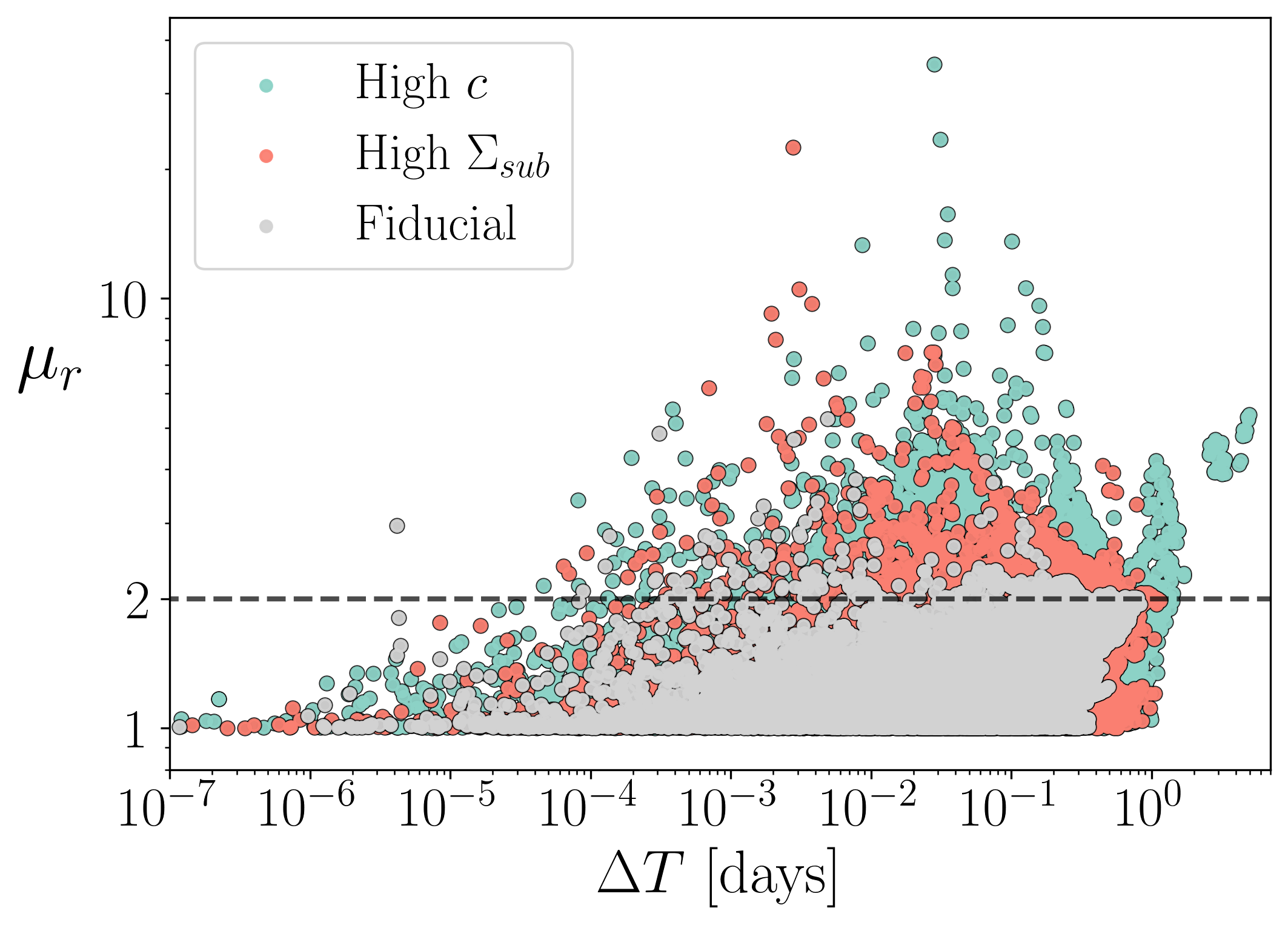}
    \caption{Relative magnifications and time delays 
    for the highest concentration $c$ and subhalo number density (parametrized by $\Sigma_{sub}$) considered in this work.The fiducial models are shown in gray.
    }
    \label{fig:mur_td_sig_con_high}
\end{figure}

Given the high number density of halos (especially in the lower mass range), we only consider subhalos near the critical curves of our main halo in this work to avoid computational limitations. These subhalos are selected through a magnification cut, considering subhalos that fall within the $|\mu| > 5$ annulus. While subhalos farther away form the critical curves (and even along the line of sight) offer exciting opportunities to learn about both subhalos and dark matter  \cite{Keeton:2008gq}, in this work we will only focus on highly magnified gravitational waves, and thus only consider subhalos in this region. This greatly reduces the number of subhalos in our simulations due to the vast majority of the total $\sim 6800$ subhalos in Fig.~\ref{fig:number_con_mass} residing very far from the critical curves of the main halo), and thus makes the computations of lensing observables (especially $I(\tau)$ and $F(f)$) feasible. We leave considering the full spatial distribution of subhalos to future work.

In order to solve for the lensing observables, we employ $\texttt{GLoW}$\footnote{\url{https://github.com/miguelzuma/GLoW_public}} \cite{Villarrubia-Rojo:2024xcj}, which is allows us not only to calculate the geometric optics observables, but the full diffraction integral to capture both the interference and diffractive effects present in GW lensing for arbitrary lens configurations.

\begin{figure}[t]
    \centering
    \includegraphics[width=\linewidth]{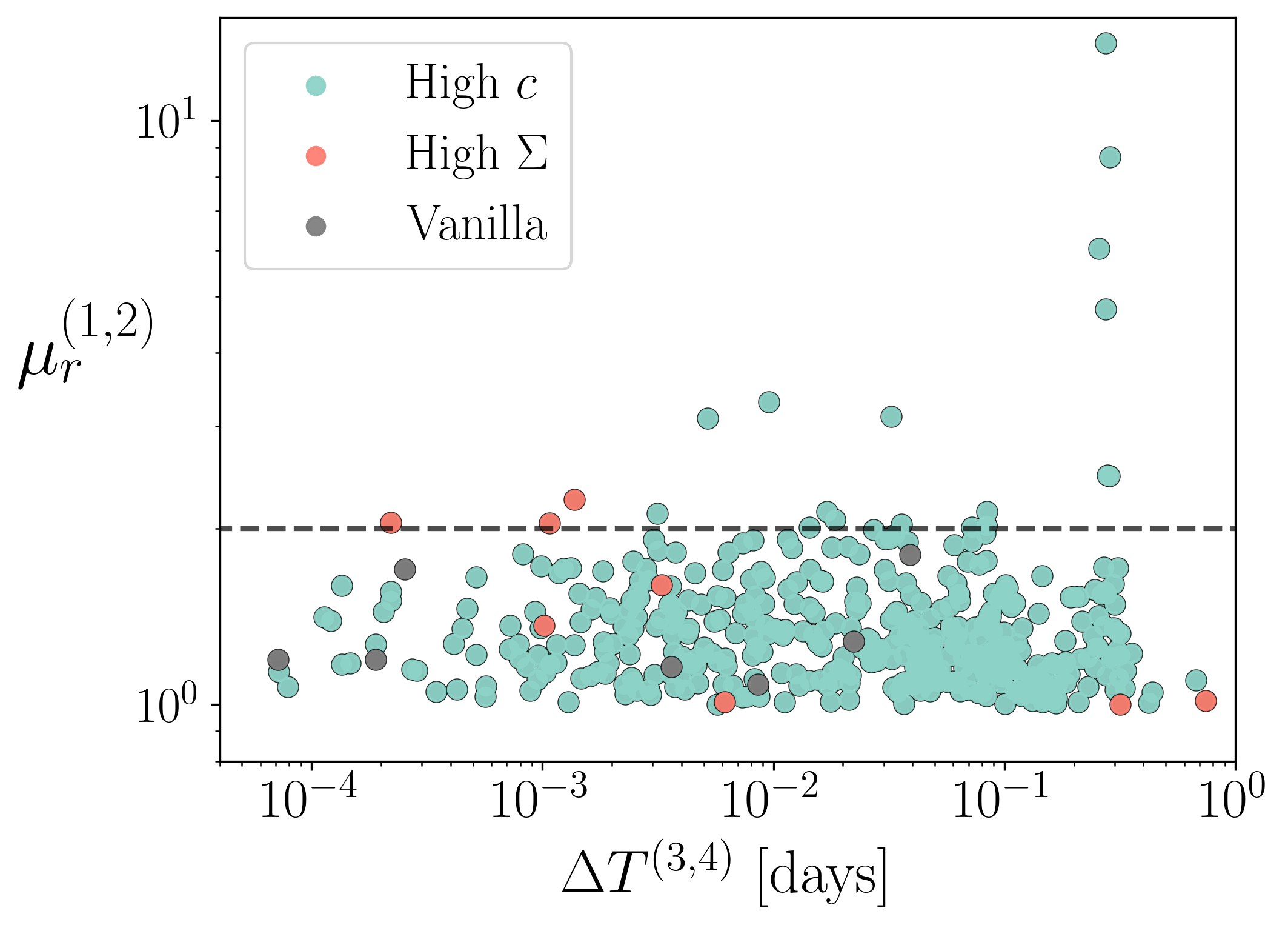}

    \caption{Relative magnifications of the brightest and second brightest images compared against the time delays of the third and fourth brightest images for the fiducial, high number density, and high concentration models. Note that the existence of the population of short time delay pairs for this image pair is due to these sources falling within higher order or nested caustics. All other source locations fall in the high \delt branch of the distributions. 
    }
    \label{fig:td_mur_highpairs}
\end{figure}

\section{Effects of Subhalos on Lensed GW Observables}
\label{sec:statistics}
\begin{table}[htb]
    \caption{Summary of the probabilities of finding a source location with $\mu_r > 2$, and generating more than 5 images ($N_{im} > 5$) for the different models in this work.}
    \label{tab:stats}
    \begin{ruledtabular}
    \begin{tabular}{lcc}
        Model & $P(\mu_r > 2)$ & $P(N_{im} > 5)$ \\
        \midrule
        Fiducial              & $9.0 \times 10^{-3}$   & $2.8 \times 10^{-4}$ \\
        $c = 3 \times c_{DJ}$ & $8.0 \times 10^{-2}$   & $6.4 \times 10^{-3}$ \\
        $c = 6 \times c_{DJ}$ & $8.7 \times 10^{-2}$   & $1.6 \times 10^{-2}$ \\
        $\Sigma_{sub} = 0.050$      & $1.0 \times 10^{-2}$   & $1.7 \times 10^{-4}$ \\
        $\Sigma_{sub} = 0.075$      & $4.0 \times 10^{-2}$   & $3.1 \times 10^{-4}$ \\
    \end{tabular}
    \end{ruledtabular}
\end{table}

In this section, we aim to quantify the fraction of the source plane in which the subhalo effects can be seen, and how this is affected by varying both the number density and concentrations of the subhalos in our simulation. 

We separate what we quantify as the effects of subhalos into two categories: the rate at which we see 4 (or more) bright images, and the rate at which we see deviations from the expected \murdelt relations (summarized in Table~\ref{tab:stats}). Due to the large separation in scales of the largest and smallest subhalo mass, we must be cautious to mitigate the effects of the resolution of our simulations. In order to do this, we first identify a region in the source plane with a minimum source magnification of $\mu_{sr} > 25$. We then densely sample the region within the boundaries set by the magnification cut in order to capture the small-scale effects of the low-mass halos. Finally, we draw source locations from the high-resolution source plane to calculate our lensing observables.

\subsection{Deviations from Universality Relations}
\label{subsec:deviations}

As outlined in \S ~\ref{subsec:caustics}, the fold and cusp caustics found in single halo lensing systems follow unique behaviors described from catastrophe theory \cite{Schneider:1992,Lo:2024wqm, Ezquiaga:2025gkd}. More specifically, a source within the main caustics (thus having an image multiplicity of greater than three) must have a relative magnification less than $\mu_r < 2$. In Fig.~\ref{fig:td_mur_clean_vs_sub}, we see that this holds in the case of the single halo (left panel), but is broken in the case of a halo populated with subhalos (right panel). These pairs of images are being generated by sources near \textit{higher order catastrophes}, which are caustics generated by the subhalos perturbing the CCs of the main halo (which will be explored in \S~\ref{subsec:highordercaustics}).
This is to be expected given the observational evidence from quasar flux anomalies \cite{Mao:1997ek,Gilman:2019nap,Nierenberg:2023tvi}, as well as excess total magnification in lensed sources near apparent cusps \cite{Aazami:2006qw}.

\subsection{Impact on Additional Images}
\label{sec:moreimages}

\begin{figure*}[t]
    \centering
    \includegraphics[width=0.475\linewidth]{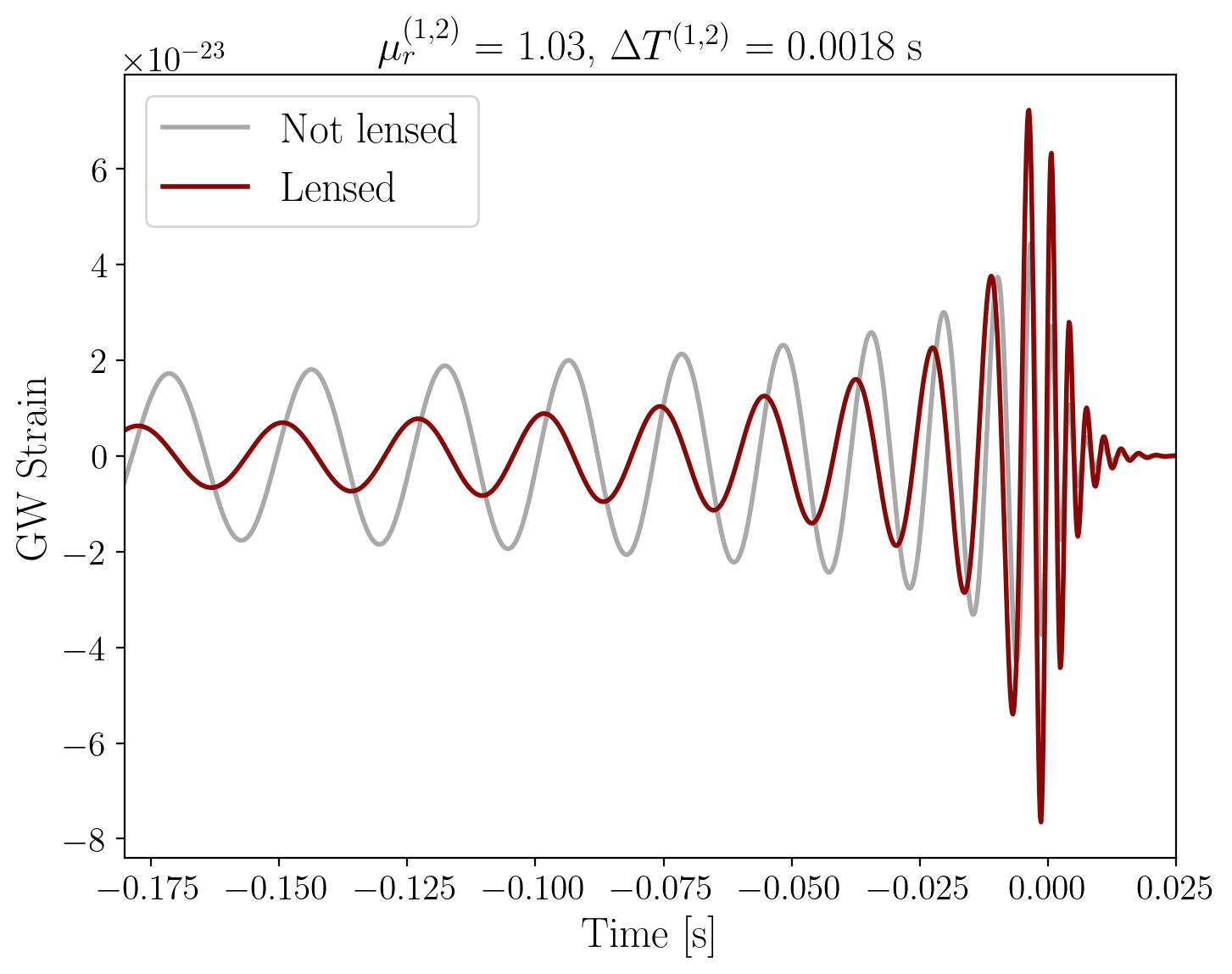}
    \includegraphics[width=0.46\linewidth]{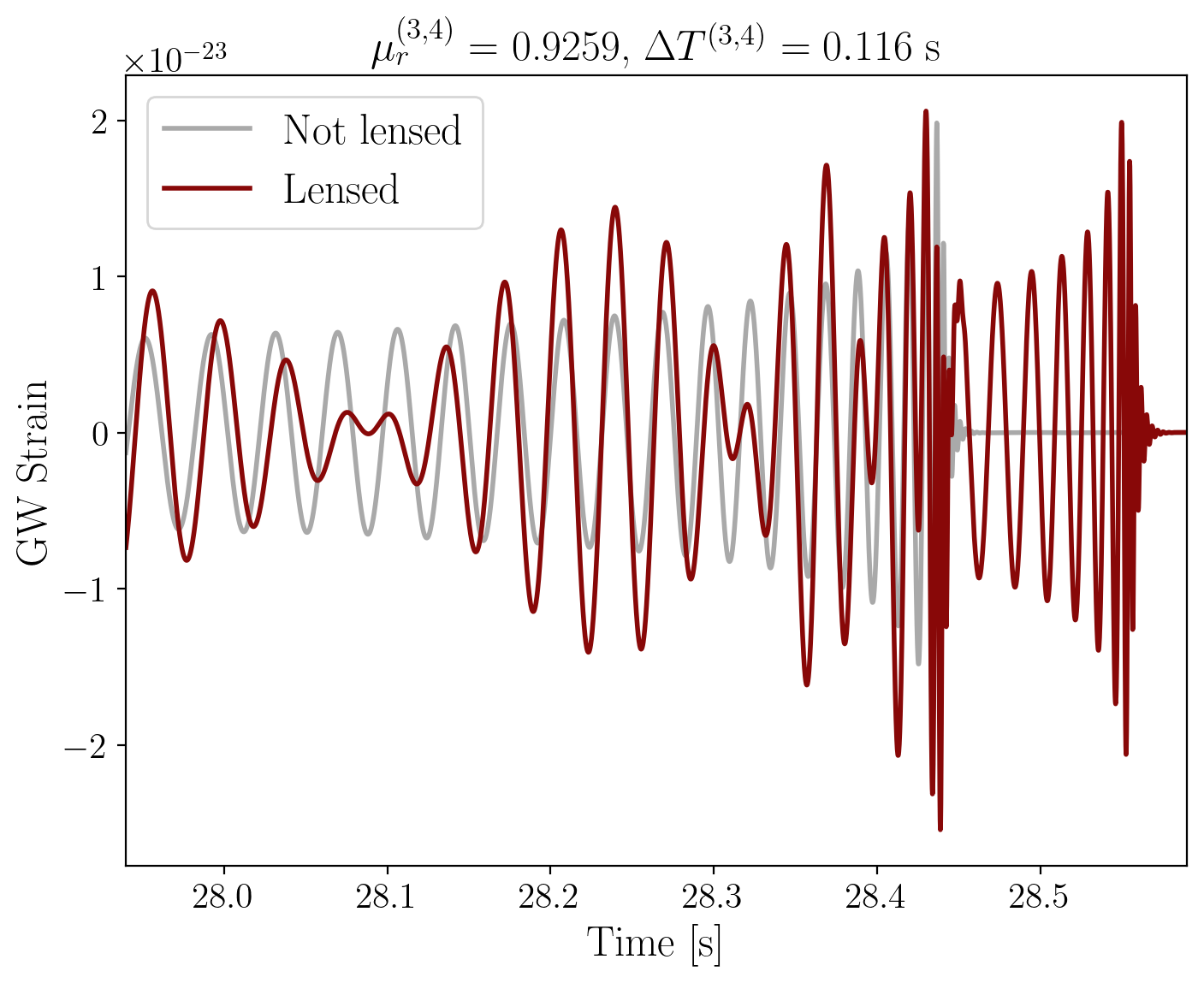}

    \caption{Lensed gravitational wave strains  
    for a source location within a higher order caustic produced by a subhalo in the example case of Fig.~\ref{fig:mag_cusp_vs_fold_subhalos}. 
    The left pane, shows the arrival of the first two images, the second shows the third and fourth images. The second and third images arrive with a delay of $\Delta T^{(2,3)} = 28.44$ s, with a relative magnification of $\mu_r^{(2,3)}=5$.  The gray lines show the (rescaled) unlensed signal aligned at the time of peak strain of the first image. 
    Note that the superscript denotes the relative arrival time of the images, and are not ordered by strain amplitude as in the rest of the text (i.e. $\mu_r^{(i,j)} = \mu_j / \mu_i$, and $\Delta T^{(i,j)} = |T_j - T_i|$.)
    }
    \label{fig:waveforms}
\end{figure*}

One of the key properties having either higher order caustics or nested caustics
caused by subhalos is the possibility of higher image multiplicities. This has been explored in detail in the context of higher lens mass systems in both the EM \cite{williams2023flashlights} as well as in gravitational waves \cite{Vujeva:2025kko}. In the context of subhalos, the occurrence of $N_{im} > 5$ is expected to be a rare occurrence due to the small strong lensing cross section of low mass subhalos. An example of this can be seen visually in Fig.~\ref{fig:mag_cusp_vs_fold_subhalos}, where there are only very small regions within the caustics of the main halo where there are additional caustic crossings due to either disconnected or higher order caustics. We report the probabilities of seeing higher image multiplicities in Table~\ref{tab:stats}. 

Despite the rarity, even if we do not see the production of additional images, we expect to see the effects of subhalos in the image properties of images arriving at later times, which should differ from those of a smooth single halo. For instance, sources within a higher order caustic are expected to produce four highly magnified images, which differs from the maximum of three highly magnified images produced by cusps (which is the highest one can produce in a single lens system). Therefore, in order to study these effects, 
we can examine the relative arrival times of the third and fourth brightest images $\Delta T^{(3,4)}$.
We see that small values of $\Delta T^{(3,4)}$ (shown in Fig.~\ref{fig:td_mur_highpairs}) correspond to source locations that either fall within higher order caustics, or nested caustics. In both cases, the total image multiplicity of these locations in the source plane are determined by how many caustic one must cross in order to reach the region of interest (where every caustic crossing corresponds to the creation of two additional images). We also see that $\Delta T^{(3,4)}$ can be short enough to cause interfering signals, opening the possibility for multiple sets of overlapping images pairs for a single source location. An explicit example of this can be seen in Fig.~\ref{fig:waveforms}, where two distinct pairs of overlapping signals coming from one source location.

\subsection{Discerning Between Concentration and Number Density}
\label{sec:n_vs_sub}

\begin{figure}
    \centering
    \includegraphics[width=\linewidth]{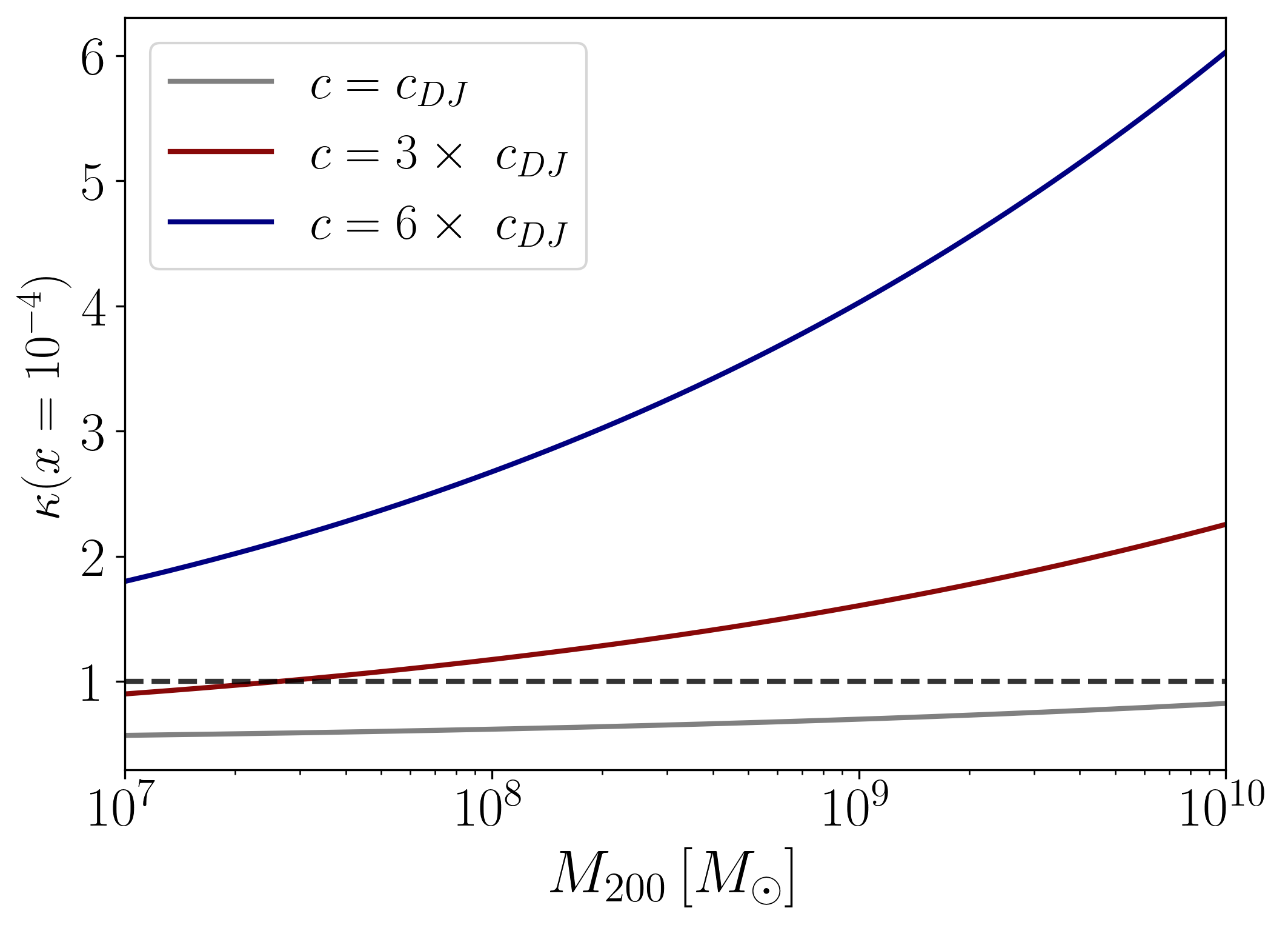}
    \caption{Central convergence of NFW halos near the critical curves of the fiducial main halo. 
    The black line corresponds to the strong lensing criterion of $\kappa=1$. 
    }
    \label{fig:maxkappa}
\end{figure}

We proceed by investigating two modifications to the subhalos in our model: their number density ($N_{sub}$), and their concentrations ($c$). The number density is modified through changing the normalization parameter $\Sigma_{sub}$ of the subhalo mass function at $10^8 M_\odot$ (see \cite{Gilman:2019nap}). We test parameters of $\Sigma_{sub} = [0.025, 0.050, 0.075]$. The concentrations are parametrized using a modified Diemer-Joyce relation \cite{Diemer:2018vmz}, where concentrations drawn from this model are rescaled by a constant factor $c_{scale} = [1, 3, 6]$. The case of $c= 3 \times c_{DJ}$ is of particular interest, as recent studies have found this to be consistent with subhalos with masses $M_{vir} \lesssim 10^9 M_\odot$ \cite{Ishiyama:2020vao} (which is approximately the maximum subhalo mass considered in this work), and has been explored in other works focusing on weakly lensed GWs in the presence of subhalos \cite{Brando:2024inp}.

We find that as we increase our two parameters, we see distinct features, particularly in the case of high concentrations. In the high $\Sigma_{sub}$ case (Fig.~\ref{fig:mur_td_sig_con_high} and Fig.~\ref{fig:mur_td_sig_con}), we see that as we increase the number of subhalos in our simulation, there are considerably more image pairs that cross our $\mu_r > 2$ threshold. This is unsurprising given that as we increase the number of subhalos near the critical curves, we increase the number of higher-order caustics generated, thus increasing the probability of seeing their effects on an image pair. 

In the high-concentration case, we generate significantly more image pairs above our threshold (substantially more than in the increased $\Sigma_{sub}$ case as summarized in Table~\ref{tab:stats}).
Additionally, embedded within \murdelt distribution we see the qualitative features of sources placed near higher order caustics (Fig.~\ref{fig:td_mur_but_swa}), as well as features akin to those of sources just outside of low mass caustics. These features arise due to the effects of nested caustics embedded within the caustics of the main halo. This feature feature is the spire-like branches of high \mur image pairs (see Fig.~\ref{fig:td_mur_invsout}) embedded in the short \delt distribution due to their low mass.

To understand the efficiency at which such caustics can be generated for NFW subhalos, we consider a typical condition for generating strong lensing, having a point in a lens mapping where $\kappa > 1$. This is because below this threshold, the mapping remains globally invertible. Above this threshold, the mapping is no longer injective due to the generation of multiple images for single source locations. We estimate the efficiency of subhalos to generate critical curves disconnected and outside of the main halo (which generate nested caustics) by computing the central convergence of the halos in close proximity to the caustics. 

\begin{figure*}[t]
    \centering
    \includegraphics[width=\linewidth]{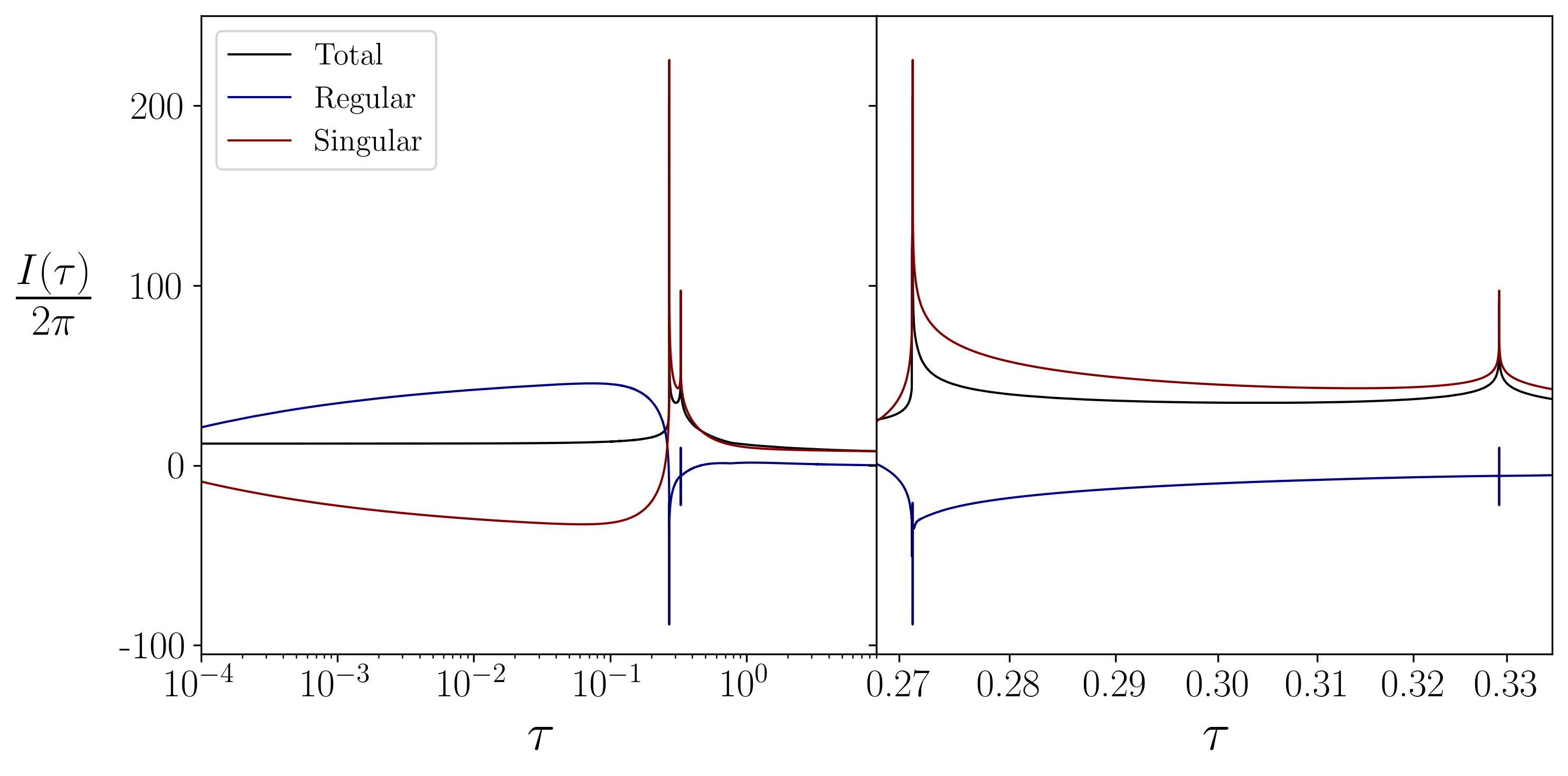}

    \caption{Time domain amplification factor $I(\tau)$ shown alongside its regular and singular components 
    (left) for the example source location generating the lensed signals in Fig.~\ref{fig:waveforms}, as well as a zoom in around the main peaks (right) to show the contributions of the regular and singular parts of $I(\tau)$ near the highly magnified images. 
    }
    \label{fig:itau_fomega_comb}
\end{figure*}

For an NFW halo, the 2D surface mass density is given by 

\begin{equation}
    \Sigma(x) = \frac{2\rho_s r_s}{x^2 - 1}f(x),
\end{equation}
where $f(x)$ is
\begin{equation}
f(x) = 
\begin{cases}
1 - \frac{2}{\sqrt{1 - x^2}} \mathrm{arctanh} \sqrt{\frac{1 - x}{1 + x}} , & x < 1 \\[10pt]
0, & x = 1 \\[10pt]
1 - \frac{2}{\sqrt{x^2 - 1}} \mathrm{arctan} \sqrt{\frac{x - 1}{1 + x}}, & x > 1
\end{cases} \ .
\end{equation}

 We approximate the central convergence of the NFW subhalos in isolation by taking the convergence at $x=10^{-4}$. 
 We show the central convergence of NFW subhalos following both the fiducial and modified $c-M$ relations for subhalos near the main CCs in Fig.~\ref{fig:maxkappa}. We make the simplifying assumption that the contribution to the total convergence coming from the main halo near the main CCs is $\kappa \approx 1/2$, and can therefore write the central convergence of the subhalos near the CCs as 
 
\begin{equation}
    \kappa_c \approx \kappa(0) + 1/2. 
\end{equation}

We see that the fiducial subhalos embedded near the CCs fall below the critical $\kappa>1$ threshold for the entire mass range, making it unlikely for them to form nested caustics, even in the case of high subhalo number densities due to the low central density of each individual subhalo. However, even small increases to the subhalo concentrations shift much lower subhalo masses above the strong lensing threshold, thus allowing for the production of significantly more nested caustics. 
This is even more exaggerated in the high concentration case, where even the lowest mass subhalos in our simulation are efficient enough to generate disconnected caustics. This most likely explains the discrepancy of the rate of seeing both greater than 5 images ($N_{im}$) and $\mu_r > 2$ in the high concentration case as opposed to the high number density case. We also find that nested caustics are more efficient at creating source plane regions where $N_{im}$, which explains why there is a much higher probability of seeing this in the high concentration case as opposed to the high number density case.
This also provides insight into the range of masses dominating the contribution to these effects,  which are found to be subhalos of masses greater than $10^8 M_\odot$ in both the fiducial and high number density simulations, whereas we see an increased contribution from lower mass halos in the high concentration simulations.

\begin{figure*}[t]
    \centering
    \includegraphics[width=0.475\linewidth]{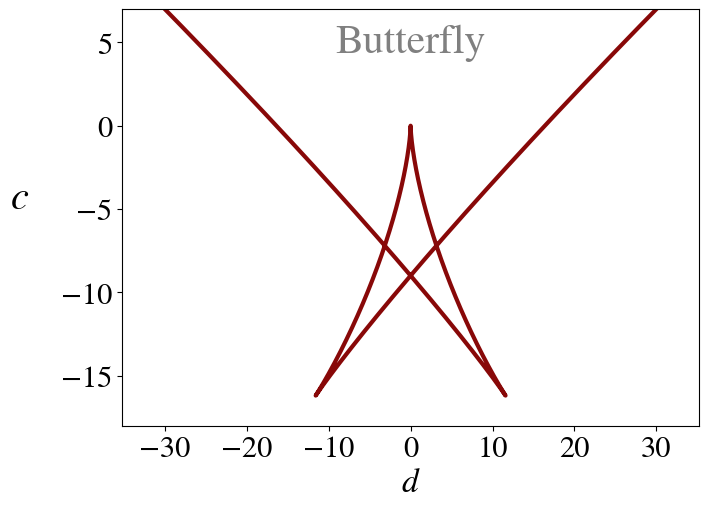}\includegraphics[width=0.475\linewidth]{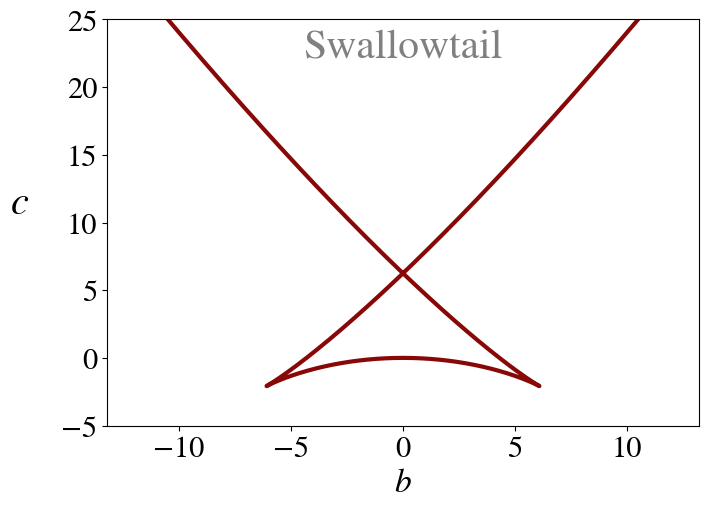}
    
    \caption{Bifurcation sets of the butterfly (left) and swallowtail (right) catastrophes. In the lensing analogy, the center of the lens is towards the top of the plot. 
    We use Arnold's parametrization~\cite{Arnold:1992gf} described in \S\ref{subsec:highordercaustics}.
    }
    \label{fig:bifurcation_swallowtail_butterfly}
\end{figure*}

\subsection{Impact on $I(\tau)$}
\label{subsec:I_tau}

To further study the implications of substructure on the lensed signals, we examine the time dependent amplification function $I(\tau)$ in Fig.~\ref{fig:itau_fomega_comb}. This calculation is enabled by creating our setup using \glow, which allows for the direct computation of these quantities. Due to the highly oscillatory nature of the integrand of the diffraction integral, computing $F(w)$ at the high $w$ corresponding to LISA or LVK signals (note that the dimensionless frequencies for these detectors for our fiducial redshifted lens mass is $w \sim 10^4 - 10^{10}$) is currently not possible for all frequencies required to reconstruct lensed signals for either detectors in our setup. Calculating these quantities at higher frequencies is left to the future work of finding a more efficient way of calculating the high frequency portion of this integral. However, we present the results for the lower frequency portion that is currently calculable as a proof of concept. 

We select the same source location used to generate the time domain waveform in Fig.~\ref{fig:waveforms}.
We see distinct structure in the regular and singular parts of $I(\tau)$ in Fig.~\ref{fig:itau_fomega_comb}, especially near the prominent peaks corresponding to the highly magnified images. Of particular interest are the highly negative portions of the regular portion of $I(\tau)$ near the peaks, indicating the existence of wave optics phenomena occurring for these images not fully captured by the singular piece.
This will correspond to distinct irregular oscillations in $F(w)$ that are characteristic of the inclusion of additional structure in the lens. Similar features can be seen in the study of wave optics phenomena in the weak lensing regime caused by dark matter subhalos \cite{Brando:2024inp}. This would create unique frequency dependent oscillations for signals in this regime that is not reproducible by diffraction effects caused by a single dark matter halo (or even a compact lens such as the point mass lenses typically used to study wave optics phenomena in lensed GWs). Whilst the impact of these oscillations might be suppressed at high frequencies where we approach GO, longer wavelength signals could still undergo interesting diffraction phenomena if there are small, compact dark matter halos along the propagation path. This motivates the push towards computing the diffraction integral at higher frequencies in order to hunt for these features in substructure-lensed signals.

\section{Higher Order Caustics and Their Observables}\label{sec:results}

In this section, we introduce some basic theory about the higher order caustics created by the subhalos perturbing the critical curves of the main halo to gain a deeper understanding of why we are seeing these deviations from the typical universality relations. We first present them in their most general form, and then build toy model examples of each caustic by placing single massive halos at a point along the critical curve that maps back to either a fold, or a cusp in the source plane. This allows us to study the most general types of perturbations to the caustics we expect to see from subhalos, in isolation.

\subsection{Higher Order Catastrophes}
\label{subsec:highordercaustics}

\begin{figure*}[t]
    \centering
    \includegraphics[width=\linewidth]{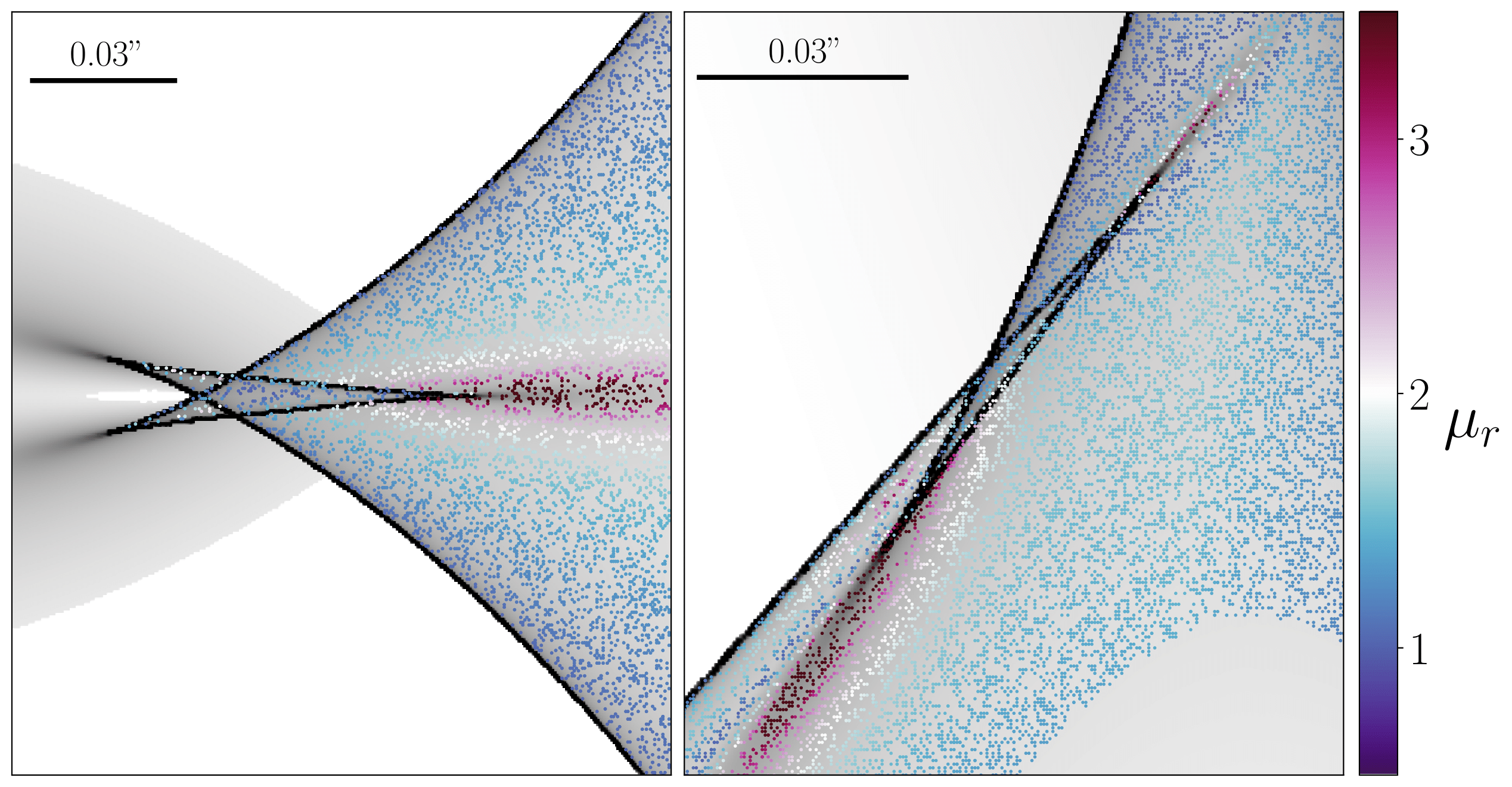}
    \caption{Source plane magnification map near the perturbed cusp (i.e. the butterfly catastrophe), and perturbed fold (the swallowtail catastrophe), with relative magnifications ($\mu_r$) of different source locations shown in color. Note that the existence of higher order caustics allows for high \mur (shown in pink), short \delt image pairs.
    }
    \label{fig:fold_pert_mur}
\end{figure*}

Starting from their most general form, we will highlight some of the qualitative features of two most common higher order catastrophes that we find in our systems, the \textit{swallowtail} and \textit{butterfly} catastrophes \footnote{Coincidentally, the swallowtail catastrophe was the subject of Salvador Dal\'i's final painting \textit{'The Swallow's Tail' (1983)}, inspired by a meeting with R\'ene Thom. Dal\'i had originally set out to paint all of Thom's catastrophes, but only completed \textit{The Swallow's Tail} before his untimely death in 1989.}.
Details about the theoretical aspects of these catastrophes (and many more) can be found in Ref.~\cite{Thom1981Modeles, BERRY1980257,Arnold:1992gf}. The implications of higher order catastrophes in lensing in the EM are explored in Ref.~\cite{Keeton:2000my, Bradac:2003hy,Keeton:2008gq, Aazami:2006qw, Aazami:2009fs,Orban:2009, Hidding:2013kka, Meena:2019pfa}, however deriving the full specific form of the generating functions discussed below using the time delay surface and computing the resulting gravitational wave observables is left to future work.

\subsubsection{The Swallowtail Catastrophe}
\label{subsubsec:swallowtail}

A swallowtail catastrophe, which has an ADE classification of $A_4$ in Arnold's notation \cite{Arnold:1992gf}, in its most general form is defined by the generating function

\begin{equation}
    V = \frac{1}{5}x^5 + \frac{1}{3}ax^3 + \frac{1}{2}bx^2 + cx,
\end{equation}\label{eq:v_sw}

where $x$ is the so called state variable, and $a,b,c$ are the control parameters. The mapping between control space (analogous to the source plane in lensing) and the generating function (analogous to the time delay surface) is given by the solutions of the gradient of the potential function $V' = 0$ (often called the equilibrium condition):

\begin{equation}
    0 = x^4 + ax^2 + bx + c.
\end{equation}\label{eq:vprim_sw}

This is analogous to solving the lens equation for a gravitational lensing system. The number of solutions to this equation corresponds to the number of local images generated by a specific catastrophe. Given its quartic nature, it can produce a maximum of 4 real roots, two real roots, or no real roots. The real roots of this equation correspond to the number of local images produced by that region of control space, whose boundaries define the bifurcation sets (in the case of lensing, the caustics). The equations describing the shape of the catastrophe in control space can be found by solving for the bifurcation set of the generating function (i.e. when $V' = V'' = 0$). 

First, we find $V''=0$, and solve for $b$ to find 

\begin{equation}
    b = -4x^3 - 2 a x, 
\end{equation}

which we can substitute into our expression for $V'$ and solve for $c$. This allows us to write the bifurcation set for the swallowtail in its parametric form

\begin{equation}
\begin{cases}
    b = -4x^3 - 2 a x \\
    c = a x^2 + 3 x^4
\end{cases}.
\end{equation}

\begin{figure*}
    \centering
    \includegraphics[width=\linewidth]{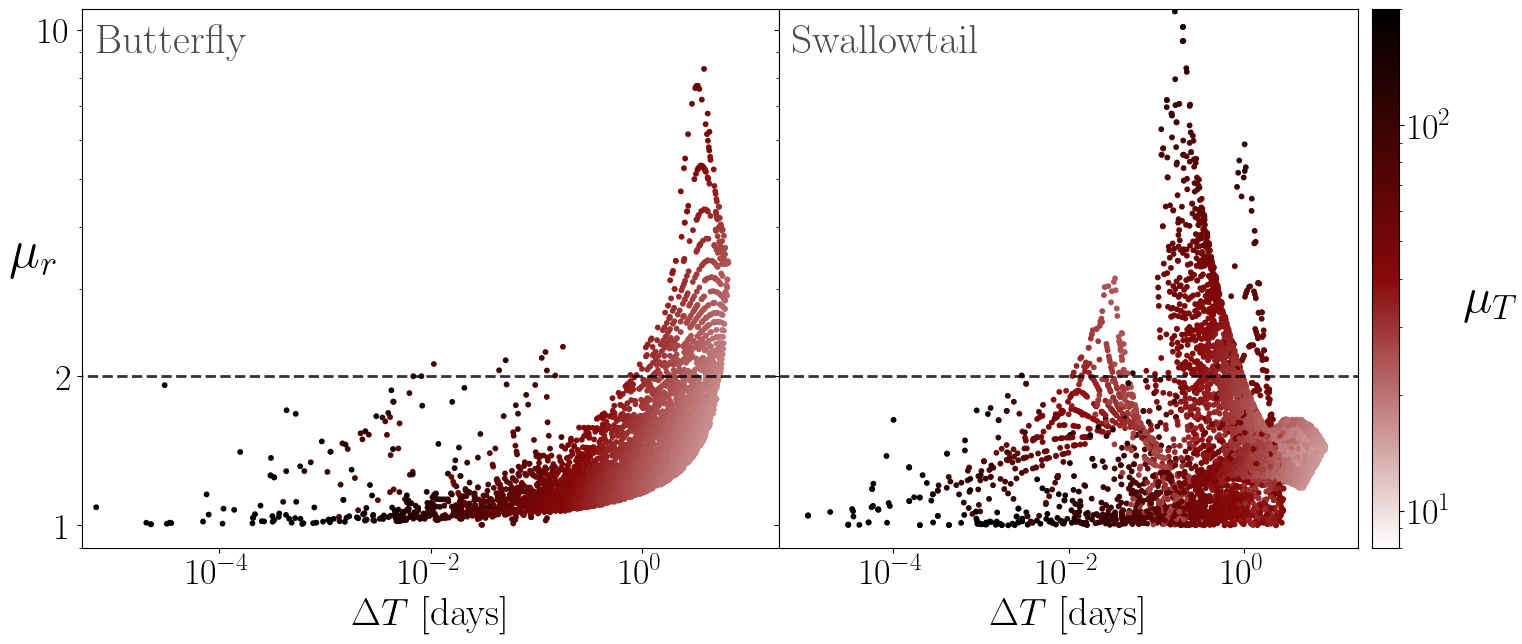}
    \caption{Relative magnifications and time delays as in     Fig.~\ref{fig:td_mur_clean_vs_sub}, but for the butterfly (left) and swallowtail (right) catastrophes shown in Fig.~\ref{fig:bifurcation_swallowtail_butterfly} and Fig.~\ref{fig:fold_pert_mur}. }
    \label{fig:td_mur_but_swa}
\end{figure*}

\subsubsection{The Butterfly Catastrophe}
\label{subsubsec:butterfly}

A butterfly catastrophe ($A_5$), in its most general form is defined by the generating function

\begin{equation}
    V = \frac{1}{6}x^6 + \frac{1}{4}ax^4 + \frac{1}{3}bx^3 + \frac{1}{2}cx^2 + dx,
\end{equation}

where we have introduced a new control parameter $d$. It's equilibrium condition $V' = 0$ is given by

\begin{equation}
    0 = x^5 + ax^3 + bx^2 + cx + d,
\end{equation}

which can produce a maximum of five real roots. Using the same approach as for the swallowtail, we can take $V''=0$, and this time solving for $c$, we find that

\begin{equation}
    c = - 5 x^4 - 3 a x^2 -2 b x.
\end{equation}

We can substitute this into the equilibrium condition and solve for $d$ to write the bifurcation set of the butterfly in its parametric form

\begin{equation}
\begin{cases}
    c = - 5 x^4 - 3 a x^2 -2 b x \\
    d =   4 x^5 + 2 a x^3 + b x^2
\end{cases}.
\end{equation}

We plot the bifurcation set (or caustics in the lensing case) for both the swallowtail and butterfly in Fig.~\ref{fig:bifurcation_swallowtail_butterfly} with a fixed value of $a=-5$ in the case of the swallowtail, and $a=-6, b=0$ for the butterfly. Note that the center of the lens is always pointed towards the top of the plot.

\subsection{Toy Model Results}
\label{subsubsec:toymodelresults}

With the insight from catastrophe theory in hand, we can proceed to building these catastrophes into our lensing setup to and solve for the observables of this toy model case.

To generate a swallowtail, we place one massive subhalo on a part the critical curve of the unperturbed main halo which maps back to a fold in the source plane. To generate the butterfly, we place one massive subhalo on a part the critical curve of the unperturbed main halo which maps back to a cusp in the source plane.
In both cases, the subhalo has the same NFW profile as the subhalos in the previous section, but with a mass of $10^{10} M_\odot$ and $c=13$ for both the fold and cusp case. Note that this choice of concentration is slightly higher than the fiducial $c-M$ relation, but is taken solely to make the effects of the subhalos clearer. The perturbed caustics can be seen in Fig.~\ref{fig:fold_pert_mur}. 

To investigate how the \murdelt relation is affected by the existence of higher-order caustics, we populate the regions of the source plane near these caustics with sources, and calculate \mur and \delt for the brightest two images. Fig.~\ref{fig:fold_pert_mur} shows that the high \mur images pairs are being produced just outside of the cusp-like feature extended towards the center of the lens. Fig.~\ref{fig:td_mur_but_swa} shows that these high \mur image pairs also have short time delays, explaining the points in Fig.~\ref{fig:td_mur_clean_vs_sub} that have both high \mur and short \delt. It is also worth noting that the cusps extended out of the main caustics (the "wings" of the butterfly), also produce high \mur image pairs, however these images have longer time delays that cannot fall within the range of time delays produced by sources inside of the main cusp and fold caustics of an unperturbed main halo, or even the perturbed main halo itself.

\section{Conclusions and Implications}
\label{sec:conclusions}

Gravitational lensing offers a unique opportunity to map dark matter structures in the Universe, and lensed GWs in particular allow us to zoom into the small scales. Their unprecedented sensitivity to short time delays --- down to milliseconds --- enables inferences of halo substructure on sub-galactic scales, competitive with other probes of small-scale structure \cite{Gilman:2019vca}. 
Although there has been a long and extensive effort to study the effects of small scale subhalos
embedded in galaxy-scale halos in EM observations, see e.g. \cite{Mao:1997ek,Dalal:2001fq}, an 
equivalent study was missing for GWs. 
In this work, we bridge this gap and study the implications for lensed GWs of dark matter subhalos.
We simulate lensing systems with subhalo populations being calculated with 
\texttt{pyHalo}~\cite{Gilman:2019nap}, and matching
recent $N$-body simulations~\cite{Diemer:2018vmz, Ishiyama:2020vao}. We test the sensitivity of GW lensing observables (number of repeated copies, relative magnifications, time delays, waveform morphologies, etc.) 
to increased subhalo number densities and concentrations. 
Our main results are summarized as follows:

\begin{itemize}
    \item 
    Substructure in the form of dark matter subhalos causes two distinct and \emph{unique} observables in order to infer their presence: 
    the breaking of universal \murdelt relations caused by the creation of higher order catastrophes, and 
    the generation
    of systems with higher image multiplicities, short time delay, and highly magnified images.
    \item The inclusion of dark matter subhalos---particularly compact, low mass subhalos---
    increases the rate of wave-optics lensing phenomena, including 
    interference and diffraction. 
    \item The rate at which subhalos influence GW observables is heavily sensitive to the $c-M$ relation, indicating a higher constraining power on dark matter models with increased concentrations at low mass subhalos.
\end{itemize}

Our results demonstrate the significance of dark matter substructures in GW lensing, opening new avenues to reveal their nature.
The last point, in particular, highlights an exciting prospect for follow-up studies to examine the effects of alternative dark matter subhalos (such as fuzzy and wave dark matter \cite{Hui:2021tkt}) that are capable of producing more compact central densities. 
This would both 
amplify the rates at which we see deviations from the \murdelt relation and higher image multiplicities, and increase the probability of seeing wave-optics phenomena. It also provides physical motivation for lensed GW parameter estimation to include more complex interfering signals, beyond the lowest order caustics~\cite{Ezquiaga:2025gkd}, as well as for GW searches to target more image pairs associated with the same lensed event.

Additionally, multi-messenger lensed events would have even further capabilities~\cite{Smith:2025axx}. They could allow us to not only combine arrival time and photometry information --- a combination that can give us insight into the physical properties of the perturber --- but also see if the DM subhalo is occupied by baryonic matter (such as a satellite galaxy). Where studies of higher order caustics are limited by spatial resolution in the EM, causing images to smear and appear similar to those from standard caustics \cite{Orban:2009}, the combination with the precision time delay information provided by GW detectors would offer unprecedented insights into the structure of such lensing systems.

As discussed in Ref.~\cite{Vujeva:2025kko}, there are challenges for the identification of these short time delay, highly magnified signals with high relative magnifications. Specifically, when inferring the lens mass solely from small numbers of lensed GW images, they could be mistaken as coming from a lower lens mass system. While this is a much more dominant effect in systems with more prevalent structures such as galaxy clusters, we find that this (albeit exciting phenomenon) could be a nuisance for inferring the total lens mass in lensed systems that do not have an EM counterpart. If not accounted for, substructures could also hinder rapid EM followups that utilize catalogs of known strong lenses such as LaStBeRu \cite{deOliveira:2025mqd} or \texttt{lenscat} \cite{Vujeva:2024scq}. 
This is because these programs aim at matching the time delay of the lensed GW signals to the lens masses in a sky localization area most likely to generate them.

The existence of unaccounted compact dark matter subhalos could also provide a hindrance to future efforts to perform time delay cosmography with lensed GWs in addition to EM observations~\cite{Gilman:2020fie}. 
Investigations in the EM have shown that not taking sub-structure into account (both in the form of sub-halos and line of sight halos) leads to an added uncertainty in the inferred value of $H_0$ that scales with the square root of the lensing volume divided by the longest time delay image used in the inference \cite{Gilman:2020fie}. 
This stresses the need to develop realistic modeling of dark matter substructures.

The discovery of the first lensed GWs is expected in the next observing run of current ground-based detectors~\cite{Ng:2017yiu,Oguri:2018muv,Xu:2021bfn,Wierda:2021upe,Smith:2022vbp}, and future space-based observatories also promise to detect more lensed signals~~\cite{LISACosmologyWorkingGroup:2022jok,Caliskan:2023zqm,Savastano:2023spl}. 
In this work, we have shown that strongly lensed GW observables are affected by dark matter subhalos, and therefore hold a key to probe their properties. As we begin to detect lensed GWs, tests such as those provided in this work will allow for unprecedented precision to the small-scale 
structures of DM subhalos, and with any luck, will shed light onto the nature of DM itself.

\begin{acknowledgements}
The authors would like to thank Juno Chan, and Guilherme Brando for their helpful comments and suggestions.
The Center of Gravity is a Center of Excellence funded by the Danish National Research Foundation under grant No. 184.
This project was supported by the research grant no. VIL37766 and no. VIL53101 from Villum Fonden, and the DNRF Chair program grant no. DNRF162 by the Danish National Research Foundation.
This project has received funding from the European Union's Horizon 2020 research and innovation programme under the Marie Sklodowska-Curie grant agreement No 101131233. 
J.M.E. is also supported by the Marie Sklodowska-Curie grant agreement No.~847523 INTERACTIONS. DG acknowledges support from the Brinson Foundation through a Brinson Prize Fellowship Grant. 
The Tycho supercomputer hosted at the SCIENCE HPC center at the University of Copenhagen was used for supporting this work.
\end{acknowledgements}

\bibliography{gw_lensing}

\appendix
\section{\murdelt Relation Inside and Outside of Caustics}
\label{ap:in_vs_out}

\begin{figure}[t]
    \centering
    \includegraphics[width=\linewidth]{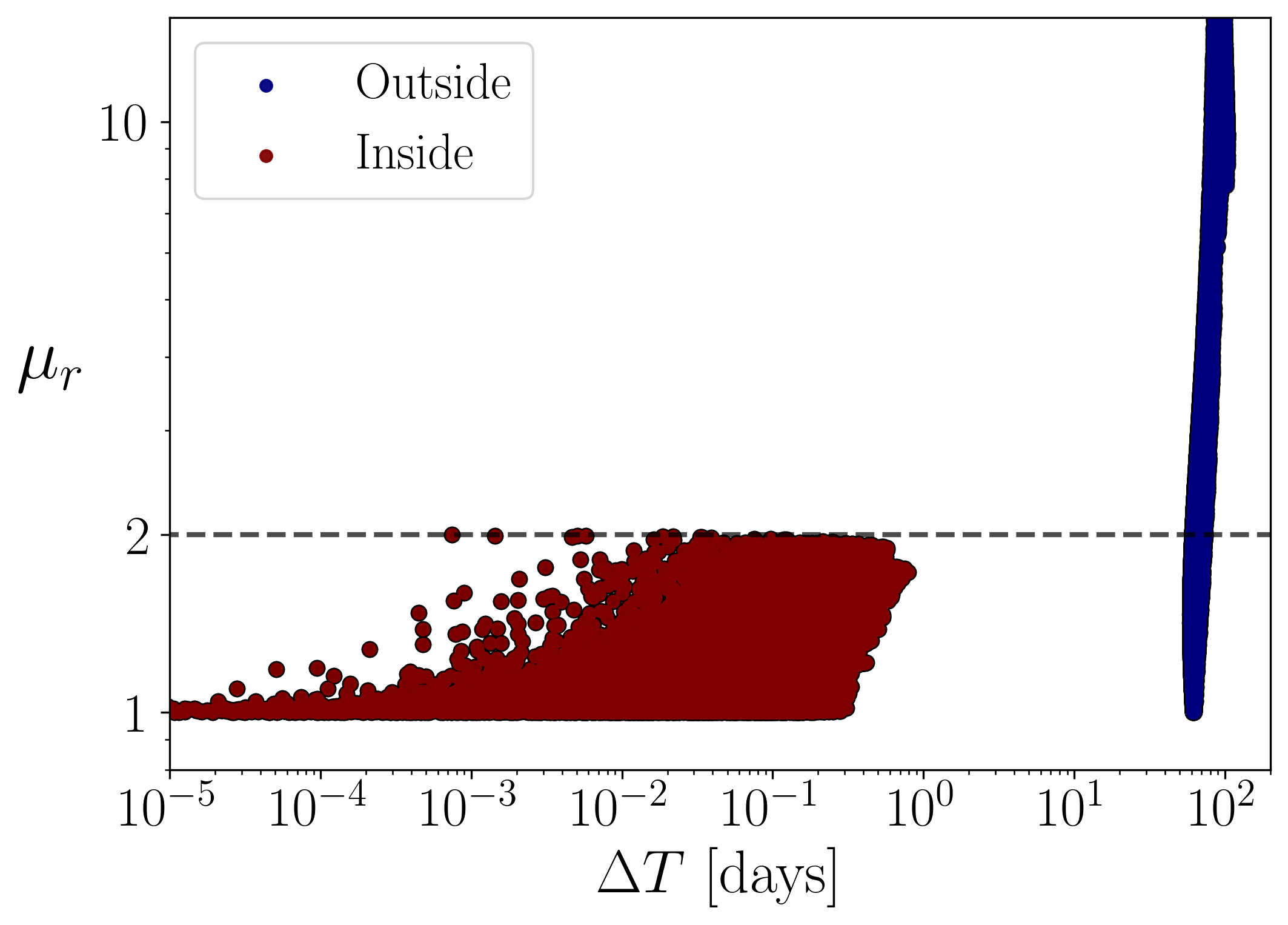}

    \caption{Relative magnification ($\mu_r$) vs time delay ($\Delta T$) for the two brightest images of sources placed near the caustics of a single elliptical single isothermal sphere halo.
    The red and blue points correspond to sources just inside and outside of the caustics respectively.
    The grey line is the maximum relative magnification factor ($\mu_r = 2$)) for sources just inside of the caustics.
    }
    \label{fig:td_mur_invsout}
\end{figure}

In this section, we will elaborate on the difference in the distribution of the \murdelt relation immediately inside and outside of the caustics. Throughout this work, we have generally been interested in the images produced by sources just inside of the main caustics of our host halo (meaning sources that fall on the side of the caustics closest to the center of the halo). This is mainly due to the images being produced in these region being highly magnified, and thus falling within the sensitivity of our current and future detectors. However, sources that lie just outside of the main caustics can still be moderately magnified by the main halo, or even highly magnified by the presence of subhalos, but the population of these images is significantly different from those of the sources just inside of the caustics. Namely, the two brightest images of a source just outside of the caustics are not subject to the same universality relations as those of sources just inside of the caustics. This can be seen explicitly in Fig.~\ref{fig:td_mur_invsout}, where the red and blue points correspond to sources just inside and outside of the caustics respectively.

The sources inside of the caustics produce the familiar highly magnified, short time delay image pairs that are of interest for this work. However, despite their proximity to the caustics, the sources just outside of them can reach high relative magnifications, but arrive with much larger time delays.

\section{Sensitivity to $c$ and $\Sigma_{sub}$}

\begin{figure*}[t]
    \centering
    \includegraphics[width=\linewidth]{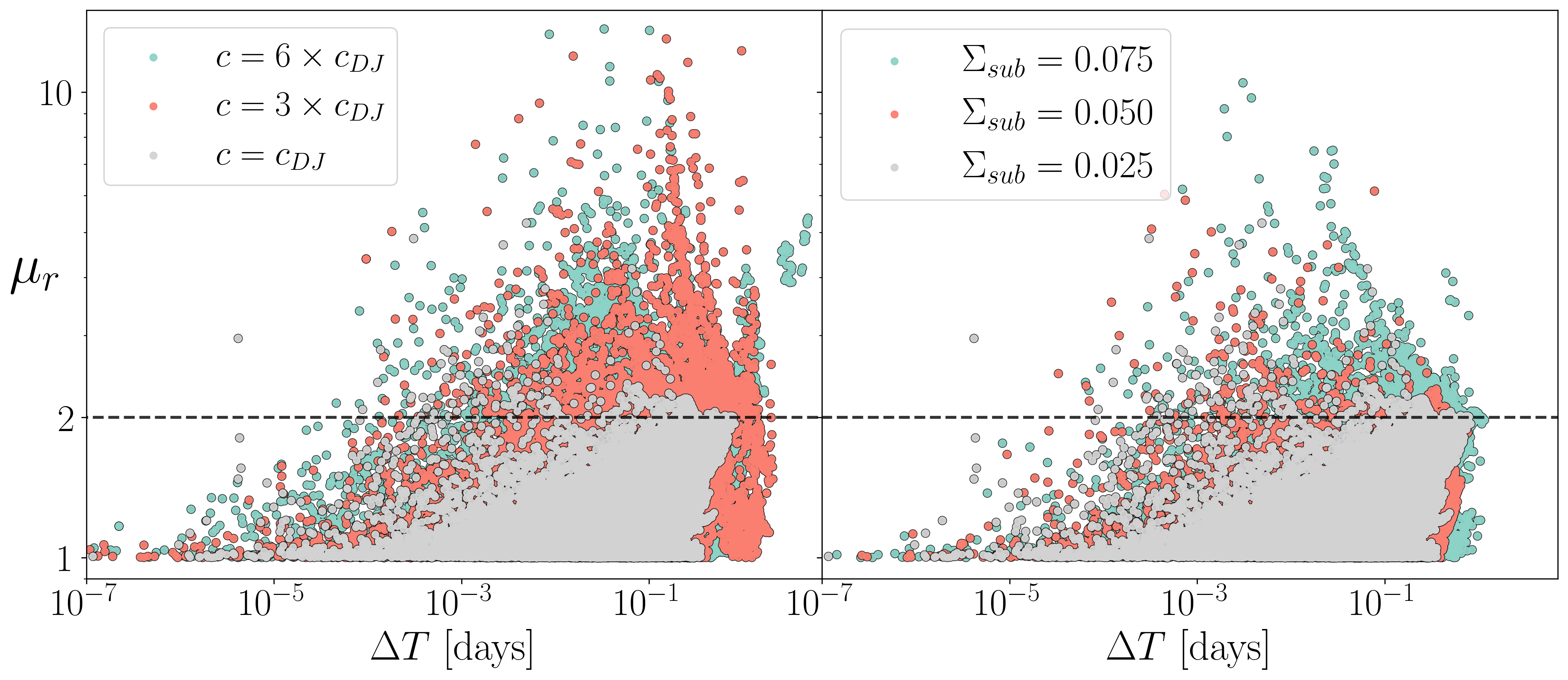}

    \caption{Relative magnifications and time delays for modified concentrations (left), and number densities (right). In both cases, the fiducial models are shown in gray. Note that $\Sigma_{sub}$ denotes the normalization of the subhalo mass function at $10^8 M_\odot$, and is not the 2D surface mass density.}
    \label{fig:mur_td_sig_con}
\end{figure*}

In this section, we present the full distributions for alternative subhalo parameters we consider in this work. Namely, we focus on changing the subhalo concentrations ($c$) and subhalo number densities (parametrized by $\Sigma_{sub}$) to both quantify the number of source locations that generate images that meet our criteria for the the images being affected by subhalos, as well as to identify qualitative differences in the \murdelt distributions.

\end{document}